\documentclass[journal]{vgtc}                
\ifpdf
  \pdfoutput=1\relax                   
  \usepackage{graphicx}                
  \DeclareGraphicsExtensions{.pdf,.png,.jpg,.jpeg} 
\else
  \usepackage{graphicx}                
  \DeclareGraphicsExtensions{.eps}     
\fi%

\graphicspath{{figures/}{pictures/}{images/}{./}} 

\usepackage{microtype}                 
\PassOptionsToPackage{warn}{textcomp}  
\usepackage{textcomp}                  
\usepackage{mathptmx}                  
\usepackage{times}                     
\usepackage{cite}                      
\usepackage{tabu}                      
\usepackage{booktabs}                  
\usepackage{enumitem}
\usepackage{bm}
\usepackage{amsfonts}
\usepackage{amsmath}
\usepackage{multirow}
\usepackage{booktabs}
\usepackage{color}

\onlineid{1696}

\vgtccategory{Research}
\vgtcpapertype{algorithm/technique}

\title{VDL-Surrogate: A View-Dependent Latent-based Model for Parameter Space Exploration of Ensemble Simulations }

\author{Neng~Shi,
        Jiayi~Xu,
        Haoyu~Li,
        Hanqi~Guo, \textit{Member,~IEEE,}
        Jonathan~Woodring,
        and~Han-Wei~Shen, \textit{Member,~IEEE}
        }
\authorfooter{
\item Neng Shi, Jiayi Xu, Haoyu Li, and Han-Wei Shen are with the Department of Computer Science and Engineering, The Ohio State University, Columbus, OH, 43210, USA. \protect\\
E-mail: \{shi.1337, xu.2205, li.8460, shen.94\}@osu.edu
\item Hanqi Guo is with the Mathematics and Computer Science Division, Argonne National Laboratory, Lemont, IL 60439, USA.\protect\\
E-mail: hguo@anl.gov
\item Jonathan Woodring is with the Applied Computer Science Group (CCS-7), Los Alamos National Laboratory, Los Alamos, NM 87544.\protect\\
Email: woodring@lanl.gov
}

\shortauthortitle{Shi \MakeLowercase{\textit{et al.}}: VDL-Surrogate: A View-Dependent Latent-based Model for Parameter Space Exploration of Ensemble Simulations}

\abstract{
We propose VDL-Surrogate, a view-dependent neural-network-latent-based surrogate model for parameter space exploration of ensemble simulations that allows high-resolution visualizations and user-specified visual mappings.
Surrogate-enabled parameter space exploration allows domain scientists to preview simulation results without having to run a large number of computationally costly simulations. 
Limited by computational resources, however, existing surrogate models may not produce previews with sufficient resolution for visualization and analysis.
To improve the efficient use of computational resources and support high-resolution exploration, we perform ray casting from different viewpoints to collect samples and produce compact latent representations. 
This latent encoding process reduces the cost of surrogate model training while maintaining the output quality. 
In the model training stage, we select viewpoints to cover the whole viewing sphere and train corresponding VDL-Surrogate models for the selected viewpoints. 
In the model inference stage, we predict the latent representations at previously selected viewpoints and decode the latent representations to data space. 
For any given viewpoint, we make interpolations over decoded data at selected viewpoints and generate visualizations with user-specified visual mappings. 
We show the effectiveness and efficiency of VDL-Surrogate in cosmological and ocean simulations with quantitative and qualitative evaluations. 
Source code is publicly available at  \url{https://github.com/trainsn/VDL-Surrogate}.
} 


\CCScatlist{ 
 \CCScat{K.6.1}{Management of Computing and Information Systems}%
{Project and People Management}{Life Cycle};
 \CCScat{K.7.m}{The Computing Profession}{Miscellaneous}{Ethics}
}

\vgtcinsertpkg

\begin{document}


\firstsection{Introduction}

\maketitle


In many scientific disciplines such as cosmology and oceanography, scientists perform ensemble simulations to analyze and explore simulation parameters. 
For example, Nyx~\cite{almgren2013nyx} and Model for Prediction Across Scales-Ocean (MPAS-Ocean)~\cite{ringler2013multi}, respectively, are models used for ensemble simulations to simulate large-scale cosmological and ocean phenomena.
By feeding different multi-dimensional input parameters, cosmologists and oceanographers can perform many simulations to observe phenomena under different conditions.
Scientific visualization techniques help scientists efficiently and intuitively analyze the similarities and differences between ensemble runs and find how different parameter settings influence the simulation outcomes.
However, running ensemble simulations with a large number of settings in the simulation parameters is prohibitively expensive and hence impractical under current computing conditions. 
For example, a model-year of climate simulation may take hours on a supercomputer~\cite{petersen2019evaluation}. 

One solution to make the simulation parameter space exploration efficient is to train a visualization surrogate to preview simulation outputs. 
Existing surrogate-model-based parameter space analyses are either data-space (e.g., NNVA~\cite{hazarika2019nnva, shi2022gnn}) or image-space (e.g., InSituNet~\cite{he2019insitunet}).
Neither method can produce satisfactory high-quality preview images, however, for the following reasons.
First, existing data-space models predict the simulation output and then visualize the predicted data. 
Limited by computing resources (e.g., GPU memory) and model efficiency, it is difficult for these surrogate models to produce high-resolution simulation output. 
Second, the image-space methods directly produce 2D images from the simulation output and use them as the training data for given viewpoints to train the visualization surrogate. 
After the surrogate models are trained, only images can be generated, and scientists cannot adjust the visual mappings, limiting the opportunity to discover and analyze previously unseen phenomena.

To address the aforementioned limitations, we propose a novel view-dependent approach to create a visualization surrogate that allows a high-quality preview of simulation output for the purpose of parameter space exploration. 
The key of our surrogate approach is to exploit view-dependent latent representations encoded from the raw data for training and prediction. First, we explain why a view-dependent strategy is necessary. 
When an appropriate resolution of visualization images is decided a priori, the visualization surrogate can be trained to produce output with the same image resolution but focus on maintaining the quality of the simulation output along the view direction. 
This strategy is similar to performing ray casting to visualize very large-scale data sets where the complexity of the underlying algorithm is transformed from $O(N^3)$ to $O(M^2 \times N)$, where $N$ is the size of data along each dimension, and $M$ is that of the image. 
When $N >> M$, there can be significant saving in the visualization algorithm.
It is to be noted that the GPU memory constraint still does not allow us to train the surrogate model successfully if we use the view-dependent approach alone based on raw data. 
Therefore, second, we use latent representations instead of the raw data along the view direction for training and prediction to further improve the model efficiency.
Specifically, we train an autoencoder to generate view-dependent latent representations. 
The latent representation helps reduce the data along the view direction, which in turn allows us to train the surrogate models successfully given limited GPU memory.
The latent representation is finally decoded to data space, making the visualization surrogate not constrained by certain visual mappings, i.e., scientists can specify any visual mappings of choice to perform parameter space exploration. 

With the view-dependent latent representation solutions, we propose VDL-Surrogate, a view-dependent neural-network-latent-based surrogate model for parameter space exploration of ensemble simulations.
Our workflow consists of three components. 
The first component is \textbf{view-dependent latent generation}.
To ensure good coverage of the domain given the predetermined image resolution, we select three view directions parallel to the three main axes. 
For a simulation output and a selected viewpoint, we perform ray casting and use a neural-network-based autoencoder called \textbf{R}ay \textbf{A}uto\textbf{E}ncoder (RAE) to encode samples along each ray with a latent representation by a new information-driven weighted $\mathcal{L}_1$ loss. 
We train three corresponding RAEs for the three selected viewpoints, and an RAE is trained given the stored simulation outputs from a few selected simulation runs. 
For other simulation runs, we use the trained RAE encoder and generate the view-dependent latent representation from the ray samples \textit{in situ}.
The second component is \textbf{offline training of our simulation surrogate}.
For a selected viewpoint, given the simulation parameters and view-dependent latent representation pairs, we train a convolutional model called \textbf{V}iew-\textbf{D}ependent \textbf{L}atent \text{Predictor} (VDL-Predictor). 
The same as RAE, three VDL-Predictors are trained for three selected viewpoints. 
The third part is \textbf{post-hoc visualization and exploration}.  
Scientists can use the trained VDL-Predictor to predict the view-dependent latent representations, decode the latent representations by RAE decoder, and perform visualization using user-specified visual mappings. 

In summary, the main contributions of this paper are: 
\setlist{nolistsep}
\begin{itemize}[noitemsep]
  \item We propose a view-dependent latent representation approach to support parameter space exploration of ensemble simulations with high-resolution visualization results and user-specified visual mappings. 
  \item We provide a comprehensive study showing the benefits of using view-dependent and latent-based methods on Nyx and MPAS-Ocean ensemble simulations. 
\end{itemize}

\begin{figure*}
  \centering
  \includegraphics[width=\linewidth]{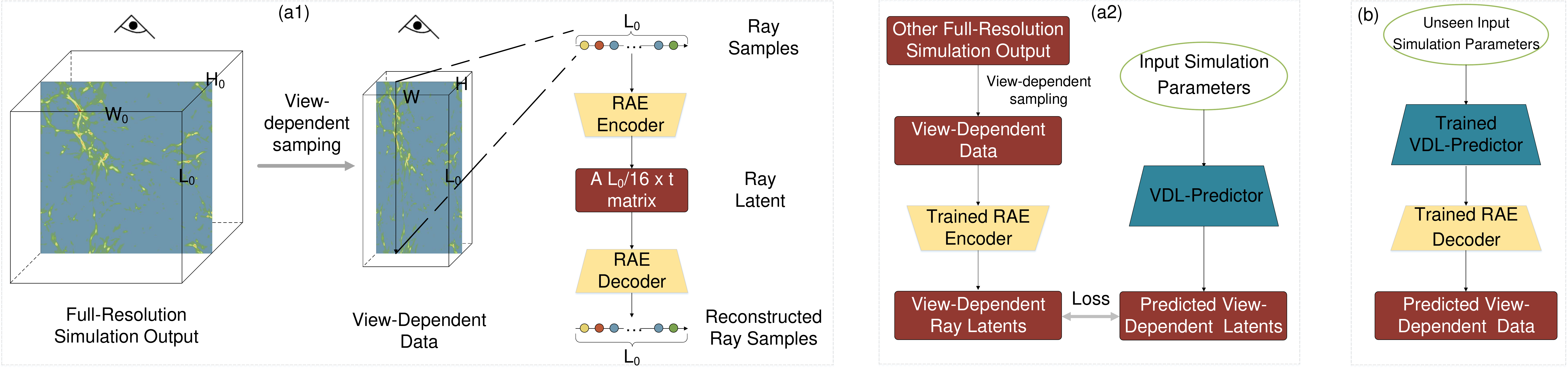}
  \caption{{ Workflow of our approach.
  (a) The model training stage. 
  (a1) View-Dependent Latent Generation.
  Given a selected viewpoint, we sample a simulation output according to the desired image resolution and collect ray samples by ray casting. 
  Next, we train a Ray AntoEncoder (RAE), which takes the ray samples as input, encodes them to compact ray latent representations, and then reconstructs them.
  (a2) VDL-Predictor Training. 
  For other simulation outputs, given the same selected viewpoint, we perform view-dependent sampling and encode the sampled simulation outputs with the trained RAE encoder in situ.
  Next, we train a VDL-Predictor, which takes the simulation parameters as input and output predicted view-dependent latent representations. 
  (b) The model inference stage. 
  Given the same selected viewpoint, we feed a new simulation parameter into the corresponding trained VDL-Predictor for a predicted view-dependent latent presentation and decode the latent representation by the trained RAE decoder to data space for visualization.} }
  \label{fig:overview}
\vspace{-0.4cm}
\end{figure*}

\section{Related Work}
In this section, we review related work in parameter space exploration and view-dependent visualization. 

\subsection{Parameter Space Exploration}
\label{subsection:param_explore}
The existing parameter space exploration work can be divided into two categories. 
The first category contains traditional methods without surrogate models, and the second is composed of surrogate-model-based methods.

Traditional parameter space exploration methods mainly focus on the collected ensemble simulation inputs and output pairs. 
Visualization researchers usually regard simulation parameters as multidimensional vectors and use techniques designed for high-dimensional data to analyze the parameter space.
These techniques contains radial plots~\cite{bruckner2010result, chen2015uncertainty, coffey2013design},  glyphs~\cite{bock2015visual}, scatter plots~\cite{orban2018drag, splechtna2015interactive, matkovic2009interactive},
parallel plots~\cite{obermaier2015visual, wang2016multi}, matrices~\cite{poco2014visual}, and line charts~\cite{biswas2016visualization}.
One significant constraint of these techniques is that they cannot explore the input parameters that have not been simulated.  

Surrogate models, including our VDL-Surrogate, are designed for parameter space exploration by predicting simulation outputs from new input parameters. 
We divide these surrogate models into (1) data-space and (2) image-space methods.
First, a data-space method predicts simulation output from unseen input parameters. 
Different techniques such as machine learning~\cite{hazarika2019nnva, alden2018using, shi2022gnn} and Gaussian process~\cite{urban2010comparison, erdal2020sampling} are applied for designing surrogate models. 
For approximating the yeast cell polarization simulation model, Hazarika et al.~\cite{hazarika2019nnva} designed a multi-layer perceptron as a surrogate model in the NNVA system. 
For better studying biological systems, Alden et al.~\cite{alden2018using} applied different machine learning approaches as surrogate models to avoid exhausting the simulation space and executing repeatedly. 
When analyzing environmental models, Erdal et al.~\cite{erdal2020sampling} employed Gaussian process emulators as surrogate models in a sampling scheme to avoid generating a large number of ensemble members. 
Due to the computing resource limitation, these works do not focus on large-scale high-resolution simulation data. 
Shi et al.~\cite{shi2022gnn} proposed GNN-Surrogate as a surrogate model for parameter space exploration of ocean simulations on unstructured grids and used an adaptive network to train the surrogate model for a simulation output with size $10^7$. 
However, the adaptive-resolution strategy is not generalizable to all the datasets. 

Second, a previous image-space method called InSituNet~\cite{he2019insitunet} visualized the simulation output in situ and generated images from different viewpoints. 
These images are rendered with several pre-defined visual mappings, which means that after they train the surrogate model, visual mappings cannot be adjusted for finding features of interest. 
Our surrogate model exploits view-dependent latent representations but is not dependent on pre-defined visual mappings to support high-resolution visualizations. 

\subsection{View-Dependent Visualization}
View-dependent methods reorganize the data samples along the rays. We organize existing related works into two paragraphs: techniques (1) not relevant and (2) relevant to simulation parameter exploration. 

Traditional view-dependent techniques focus on speeding up volume rendering and reducing data size. 
Mueller et al.~\cite{mueller1999ibr} approximated the volume rendering process by slab pre-computing at every sampled viewpoint.
Volumetric depth images are pre-computed from selected viewpoints and can be rendered with arbitrary camera configurations with low overhead~\cite{frey2013explorable, fernandes2014space}. 
An image database called ``Cinema'' was created on a large-scale ocean simulation dataset, and photos were taken from different viewpoints so that oceanographers could visualize the dataset from the images stored~\cite{ahrens2014image, ahrens2014situ, o2016cinema}.
However, these approaches do not support transfer function exploration, limiting the user's analysis flexibility. 
Tikhonova et al.~\cite{tikhonova2010explorable, tikhonova2010exploratory, tikhonova2010visualization} save image slices that can be used when users want to explore new transfer functions. 
Wang et al.~\cite{wang2018image, wang2019ray} summarized data by exploiting per-ray distributions to enable transfer function exploration.

To explore the parameter space of ensemble simulations, InSituNet~\cite{he2019insitunet} is an image-based view-dependent method visualizing the simulation outputs from different viewpoints in situ.
The major limitation of InSituNet is the inability to control visual mappings after training because their training data depend on pre-defined visual mappings. 
In our surrogate model, the view-dependent latent representation is visual-mapping-independent, enabling scientists to specify visual mappings for the simulation output preview in post-hoc analysis. 

\section{Overview}
\label{section:overview}

Our goal is to support parameter space exploration and visualization of ensemble simulations.
The parameter space exploration and visualization tasks are achieved with a visualization surrogate model that takes simulation parameters as input. 
To support simulation output visualization with high-resolution previews and user-specified visual mappings, the surrogate undergoes training and makes predictions based on view-dependent latent representations. 

\begin{figure}
  \centering
  \includegraphics[width=\linewidth]{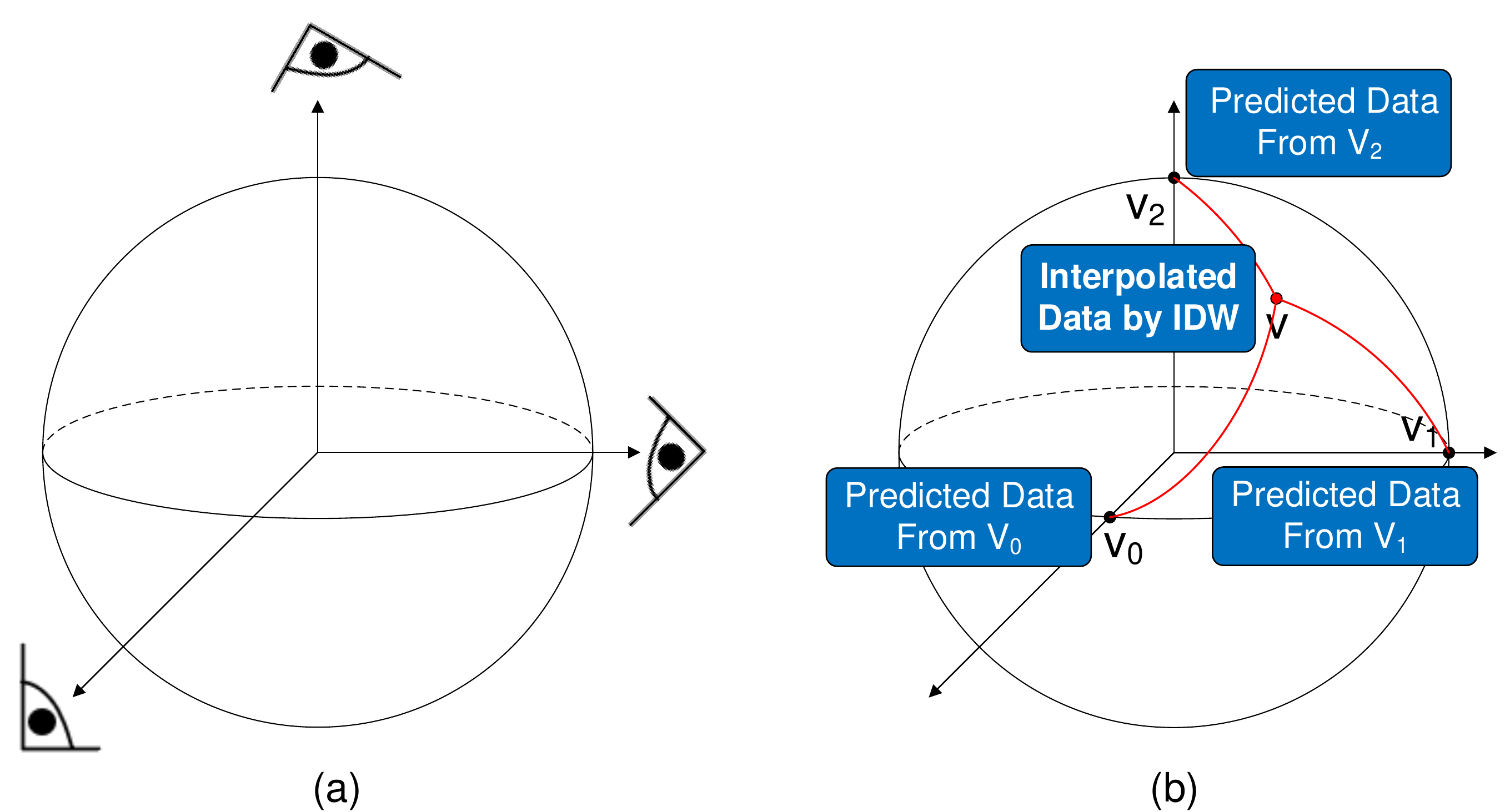}
  \caption{{Viewpoint selection and view-dependent inference.
  (a) In the model training stage, we select three view directions parallel to three main axes.
  (b) In the model inference stage, for any viewpoint $v$, the predicted simulation output visualization is from the interpolation of predicted view-dependent data at selected viewpoints $v_0$, $v_1$, and $v_2$ by inverse viewpoint distance weighting.}}
  \label{fig:view-dependence}
\vspace{-0.4cm}
\end{figure}

Figure~\ref{fig:overview} shows the workflow to construct VDL-Surrogate. 
{Note that VDL-Surrogate is composed of two major networks: \textbf{R}ay \textbf{A}uto\textbf{E}ncoder (RAE) and \textbf{V}iew-\textbf{D}ependent \textbf{L}atent \text{Predictor} (VDL-Predictor), and Figure~\ref{fig:overview}(a) shows the training process of these two networks.
Before training, we first randomly choose a small number of parameter settings to run the simulations and store the full-resolution simulation outputs to the disk. 
Considering the GPU memory constraint, in our model training stage, it is challenging to hold the intermediate feature maps of high-resolution data in the GPU memory if we use the raw data for training.
Therefore, as illustrated in Figure~\ref{fig:overview}(a1), for a saved full-resolution simulation output with size $W_0 \times H_0 \times L_0$, given a viewing direction, we sample the simulation output across the image plane to obtain view-dependent data with size $W \times H \times L_0$ according to the desired image resolution chosen for the visualization surrogate via interpolation. 
Next, with the view-dependent data, starting from each pixel location on the image plane, along the view direction, we cast rays parallel to the view axis, collect samples, and train an RAE (Section~\ref{section:vdl}) to encode a ray to a compact latent representation.
Note that we maintain the quality of the simulation output along the view direction, i.e., we keep the length of ray samples at the original resolution $L_0$.
As shown in Figure~\ref{fig:overview}(a2), for other full-resolution simulation outputs, given the selected viewpoint, we sample the simulation output according to the desired image resolution and encode the sampled simulation outputs \textit{in situ}, in order to generate the view-dependent latent representations that are used for supervising the downstream training. 
With the pair of simulation parameters and view-dependent latent representations, for the selected viewpoint, we train a convolutional model VDL-Predictor to learn the mapping from the input simulation parameters to view-dependent latent representations (Section~\ref{section:network}).
As illustrated in Figure~\ref{fig:view-dependence}(a), we select three view directions parallel to three main axes and train an RAE and VDL-Predictor for each viewing direction.}

In the inference stage, as illustrated in Figure~\ref{fig:overview}(b), for a selected viewpoint, a new simulation parameter is first fed into the corresponding trained VDL-Predictor, which will generate a predicted view-dependent latent representation.
Then this output latent representation is decoded by the RAE decoder to produce the view-dependent data for visualization.
As shown in Figure~\ref{fig:view-dependence}(b), given any viewpoint, the predicted simulation output is obtained from the interpolation of predicted outputs at selected viewpoints by inverse viewpoint distance weighting (Section ~\ref{section:inference}). 

In summary, VDL-Surrogate neural network consists of two sub-networks: (1) RAE (illustrated in Figure~\ref{fig:convae}) and (2) VDL-Predictor (illustrated in Figure~\ref{fig:vdl_pred}), which will be explained in detail as follows. 

\section{View-Dependent Latent Generation}
\label{section:vdl}

This section describes how to generate view-dependent latent representations that are later used for the downstream predictor training. 
This step helps reduce the computation and memory cost of the predictor training. 
Figure~\ref{fig:overview}(a1) illustrates the view-dependent latent representation generation process.
Given a selected viewpoint, we sample the simulation output by casting rays parallel to the axis (black line segment) from the image plane perpendicular to the view direction to collect ray samples.
Then we train a 1D convolutional autoencoder called Ray AutoEncoder (RAE) to encode and reconstruct the ray samples.
{We choose 1D convolution instead of simpler design choices such as MLPs because 1D convolutional layers help retain the input ray's sequential (spatial) structure.
Therefore, in the following VDL-Predictor, the image-wise ray latent representation also contains the ``depth'' dimension and is a good prediction target for the 3D convolutional VDL-Predictor. } 

\begin{figure}
  \centering
  \includegraphics[width=\linewidth]{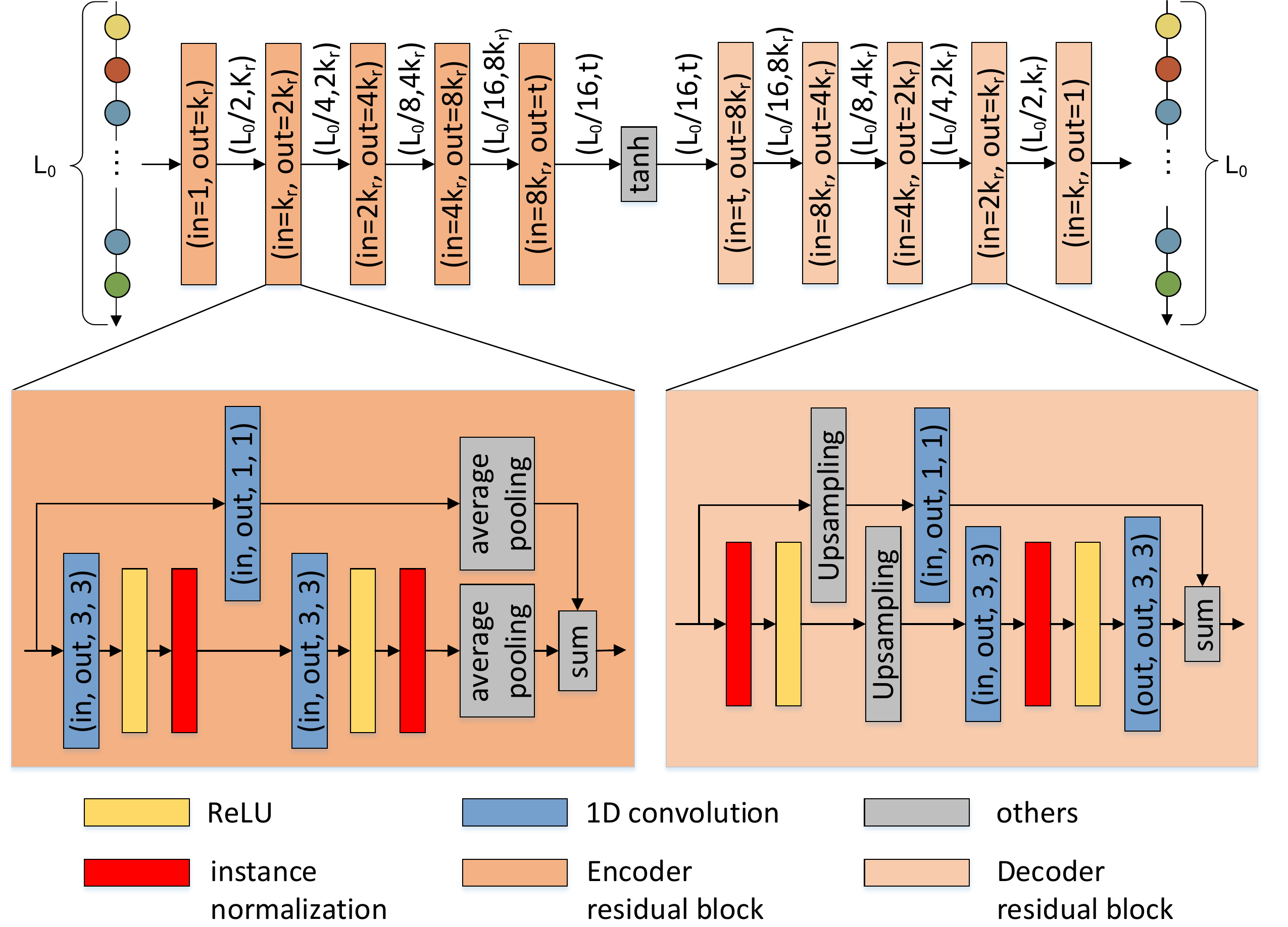}
  \caption{{The architecture of RAE, whose encoder encodes ray samples to latent representation and decoder decodes the latent representation to reconstructed ray samples. 
  The network is comprised of a series of residual blocks, and a hyper-parameter $k_r$ controls the number of network parameters. }
  }
  \label{fig:convae}
\vspace{-0.6cm}
\end{figure}

Figure~\ref{fig:convae} shows the architecture of RAE, which takes ray samples as input and performs reconstruction.
A forward pass in RAE is composed of encoding and decoding.
{In the encoding stage, a ray with $L$ samples is fed through the encoder residual blocks and a $tanh$ activation to generate a ray latent representation of the size $L_0/16 \times t$ in the range $[-1, 1]$, where $t$ is a hyper-parameter controlling the channel size of the latent representation. }
The ray latent representation captures the information of samples along the ray, which can be used for supervising the downstream predictor training.
In the decoding stage, five decoder residual blocks perform super-resolution and transform the ray latent representation into reconstructed ray samples.
We use a constant channel multiplier $k_r$ in the RAE structure to control the number of network parameters in the intermediate layers.
We apply rectified linear unit (ReLU)~\cite{nair2010rectified} as the activation function in all layers.

In the encoder residual block, we first feed the feature map into two 1D convolution layers with kernel size 3. 
Next, we perform average pooling to down-sample the feature map.  
Finally, the original feature map is added to the output, and the result is sent to the next residual block. 
The decoder residual blocks are similar to the encoder residual blocks, except we replace average pooling with nearest neighbor up-sampling. 
In RAE, Instance Normalization~\cite{ulyanov2016instance} is used for improving network convergence speed.

\begin{figure}
  \centering
  \includegraphics[width=\linewidth]{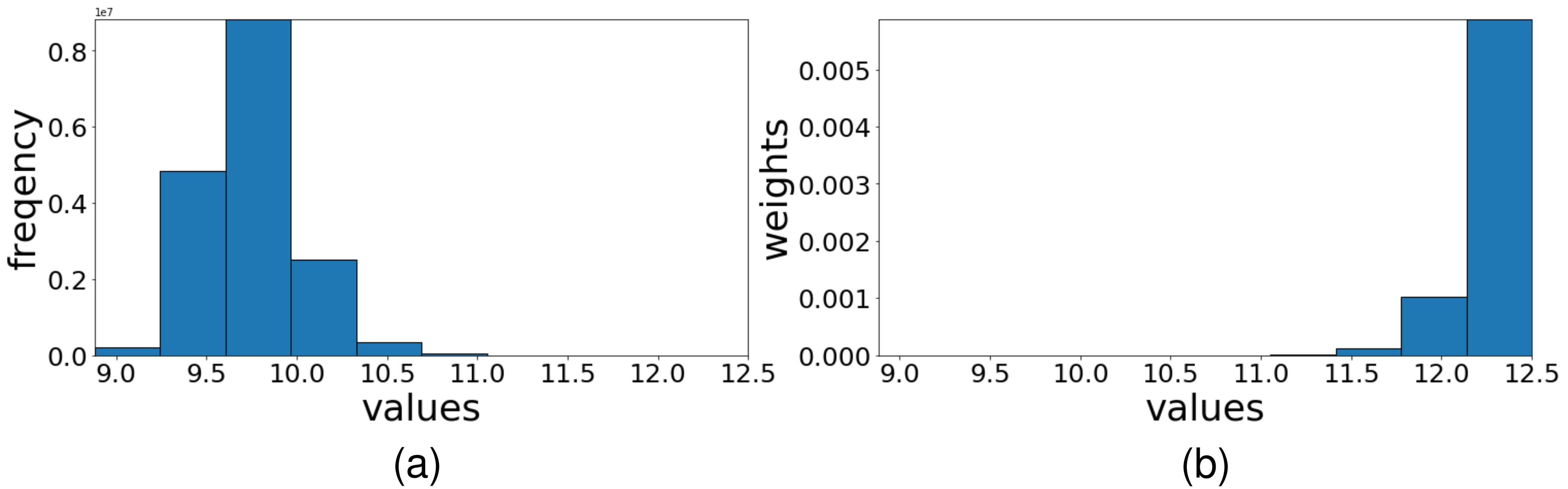}
  \caption{Illustration for the design of our information-driven weighted $\mathcal{L}_1$ loss. 
  (a) Histogram showing the frequency of data values.
  (b) Histogram showing the weights for different data values, which is the inverse of the frequency. }
  \label{fig:histogram}
\vspace{-0.4cm}
\end{figure}

Inspired by previous information-driven sampling~\cite{biswas2018situ}, we train RAE by an \textit{information-driven weighted $\mathcal{L}_1$ loss} to handle the imbalanced samples in rays. 
For example, as shown in Figure~\ref{fig:histogram}(a), in a Nyx cosmological simulation output, samples with high-density values are few but important. 
Simply applying $\mathcal{L}_1$ loss would make RAE ignore those samples.
To avoid that, we calculate the distribution of samples by a histogram in a training batch and use the inverse of frequency as the weight in the loss function, illustrated in Figure~\ref{fig:histogram}(b). 

We use two additional techniques to improve the training efficiency and stability.  
First, we train the RAE with mixed precision~\cite{micikevicius2018mixed} which reduces the training time and memory cost with minimal impact. 
Second, spectral normalization~\cite{miyato2018spectral} is applied to stabilize the RAE training. 

After the RAE training finishes, for other simulation runs, for the same viewpoint, after sampling the simulation output according to the desired image resolution, we encode the view-dependent simulation output with the trained RAE encoder \textit{in situ} to generate compact view-dependent latent representations.
The view-dependent latent representations are used for downstream VDL-Predictor training. 

\section{VDL-Predictor}
\label{section:network}

\begin{figure}
  \centering
  \includegraphics[width=\linewidth]{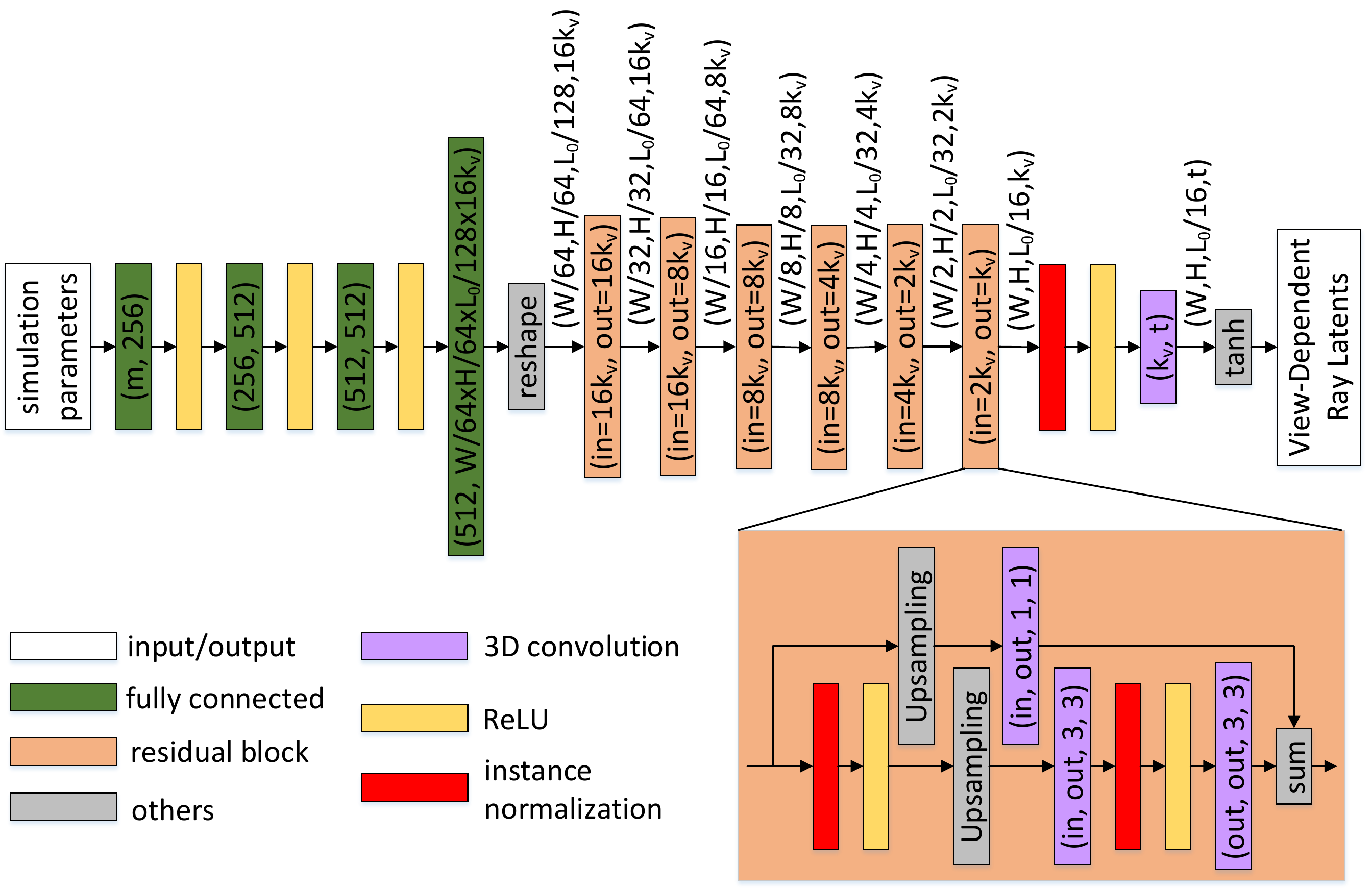}
  \caption{{The architecture of VDL-Predictor, which generates a 1D vector given input simulation parameters and maps the vector to an output predicted viewpoint-dependent latent representation. 
  One hyper-parameter $k_v$ is used to control the network size.  }}
  \label{fig:vdl_pred}
\vspace{-0.4cm}
\end{figure}

{Given an input simulation parameter setting, we predict the corresponding view-dependent latent representations for different viewpoints.} 
We design a convolutional model called \textbf{V}iew-\textbf{D}ependent \textbf{L}atent \textbf{Predictor} (VDL-Predictor) for each viewpoint to accomplish this goal.  

Figure~\ref{fig:vdl_pred} shows the architecture of VDL-Predictor, which takes a simulation parameter setting as input and outputs a view-dependent latent representation. 
There are three steps in a VDL-Predictor forward pass. 
{Our generated view-dependent latent representation is a tensor of the size $W \times H \times L_0/16 \times t$, where $W \times H$ indicates the visualization image resolution and $L_0/16 \times t$ is the latent representation shape for a single ray.  
First, we feed the input parameter setting into a fully connected layer to generate a 1D vector of size $W/64 \times H/64 \times L_0/128 \times 16k_v$, where the hyper-parameter $k_v$ is used to control the size of the network and generated vectors, similar to previously introduced $k_r$ in RAE.
Second, we reshape the vector to a tensor of the size $W/64 \times H/64 \times L_0/128 \times 16k_v$.}
Third, we use the subsequent residual blocks to perform super-resolution until reaching the desired resolution. 
Rectified linear unit (ReLU) is applied as
the activation function in all layers except the last output layer.
In the last layer, we use the $tanh$ activation function instead to normalize the output values into $[-1,1]$.
The residual block in VDL-Predictor is similar to the one in the RAE decoder.
The difference is that we replace 1D convolution kernels with 3D convolution kernels. 

In the training process, we iteratively update the parameters in VDL-Predictor using gradient descent to minimize the $\mathcal{L}_1$ Loss. 
The same as RAE training, we apply spectrum normalization~\cite{miyato2018spectral} and mixed precision~\cite{micikevicius2018mixed} to improve the VDL-Predictor training stability and efficiency, respectively.  

\section{Inference Process and Sensitivity Analysis}
\label{section:inference}
In the inference stage, we first predict the view-dependent latent representations and decode the latent representations to the data space at the three selected viewpoints.  
As illustrated in Figure~\ref{fig:view-dependence}(b), for a selected viewpoint, we first feed the input parameters into the trained VDL-Predictor and generate the predicted view-dependent latent representation.
Second, the predicted view-dependent latent representation is sent to the trained RAE decoder and then decoded back to the data space. 
The inference process is the same for the other two selected viewpoints.

Next, to visualize the predicted simulation output for a given parameter setting, our goal is to perform ray casting from any viewpoint to generate the visualization. 
For an arbitrary viewpoint $v$, we cast rays to sample the predicted viewpoint-dependent data generated from previously selected viewpoints. 
{In order to make the final visualization more realistic (size varies inversely with distance), we cast rays in the perspective mode.} 
At one ray sample location, we average all the sampled values in different predicted viewpoint-dependent data by inverse viewpoint distance weighting such that the viewpoint-dependent sampled value from a closer viewpoint should contribute more than that from a more distant viewpoint:
\begin{equation}
\hat{s} = \dfrac{\sum_{i=0}^3{\hat{s_i} \cdot q_i}}
{\sum_{i=0}^3{q_i}}, 
\end{equation} 
where $\hat{s_i}$ is a sampled value {via trilinear interpolation} in the viewpoint-dependent predicted data at selected viewpoint $v_i$, $\hat{s}$ is the weighted average of $N_v$ sampled values, 
{\begin{equation}\nonumber
q_i = \dfrac{1}{min(d(v, v_i), d(v, sym(v_i))},
\end{equation}
and $d(v, v_i)$ is the great-circle distance between viewpoint $v$ and viewpoint $v_i$, $sym(v_i)$ is the symmetrical viewpoint of $v_i$. }

Inspired by previous works~\cite{berger2018generative, he2019insitunet, shi2022gnn}, we perform sensitivity analysis on simulation parameters by utilizing the differentiability of both the RAE and VDL-Predictor.
{Specifically, for one selected viewpoint, we aggregate the predicted view-dependent data to a scalar value by $\mathcal{L}_1$ norm and calculate the derivative of that scalar value to the selected parameter.
Three derivatives are calculated from three selected viewpoints. 
We used the average of three derivatives' absolute values as the indicator for one parameter’s sensitivity because it illustrates how the predicted simulation output changes as the input parameter changes.}
We fix other input simulation parameters during one selected parameter analysis, and a uniform sampling is performed in the selected parameter’s value range.
For one sampled parameter, we first perform a VDL-Predictor forward pass and RAE decoding, and then a backward propagation is conducted for the sensitivity value. 
We repeat the process for three trained RAE and VDL-Predictors from three selected viewpoints and compute the average of three absolute derivatives as the final sensitivity value.
A line chart is applied for visualizing the list of sensitivity values.

\section{Results}
VDL-Surrogate is evaluated using two ensemble simulation datasets on cosmology and ocean  (Section~\ref{subsection:datasets}) from four perspectives.
First, we give implementation details and analyze the performance (Section~\ref{subsection:implementation}).
Second, we list the metrics used to evaluate our VDL-Surrogate (Section~\ref{subsection:metric}). 
{Third, we show the motivation for using view-dependent representations from three viewing directions and give the supporting experiment (Section~\ref{subsection:multiple_view}). }
Fourth, we compare VDL-Surrogate with baseline methods (Section~\ref{subsection:baseline}).
Fifth, we perform case studies of parameter space exploration and analysis (Section~\ref{subsection:case}). 
Moreover, The ablation studies we perform to show the necessity of different components in our model can be found in the supplementary material. 

\subsection{Ensemble Simulations}
\label{subsection:datasets}

\begin{table*} \footnotesize
\caption{Datasets and Performance. 
$k_r$ and $k_v$, respectively, control the size of RAE and VDL-Predictor for datasets with different complexities.
$t_sim$, $t_{RAE\_tr}$, and $t_{VDLP\_tr}$, are timings for running ensemble simulations, training RAE, and training VDL-Predictor, respectively; $t_{RAE\_ec}$ and $t_{VDLP\_fp} + t_{RAE\_dc}$, respectively, are the timings for an encoding of RAE and a forward pass of VDL-Predictor together with a decoding of RAE. }
	\centering
	\begin{tabular}{c|c|c|c|c|c|c|c|c}
		\multirow{2}*{Simulation}  & \multirow{2}*{Resolution}  &  \multicolumn{2}{c|}{$P_{sim}$} & \multirow{2}*{$k_r$} & \multirow{2}*{$k_v$} & \multicolumn{3}{c}{Size (GB)} \\ \cline{3-4} \cline{7-9} 
		~ & ~ & Name & Number & ~ & ~  & Raw  & RAE & VDL-Predictor  \\ \hline
		Nyx & $512 \times 512 \times 512$ & $OmM, OmB, h$ & 130 & 64 & 64 & 65.0 & $3 \times 0.013$ & $3 \times 0.66$ \\ \hline 
		MPAS-Ocean & $1536 \times 768 \times 768$  & $BwsA, CbrN, GM, HV$ & 100 & 64 & 24 & 337.5 & $3 \times 0.013$ & $3 \times 0.21$ \\ \midrule[2pt]
		\multirow{2}*{Simulation}  & \multicolumn{6}{c}{Performance} \\ \cline{2-9}
		~ & $t_{sim}$ (hr) & $t_{RAE\_tr}$ (hr) & $t_{VDLP\_tr}$ (hr) &\multicolumn{2}{c|}{$t_{RAE\_ec}$ (s)} & \multicolumn{3}{c}{$t_{VDLP\_fp} + t_{RAE\_dc}$ (s)}    \\ \hline
		Nyx & 139.8 & 24.0 & 28.7 & \multicolumn{2}{c|}{10.1} & \multicolumn{3}{c}{9.2}   \\ \hline
		MPAS-Ocean & 82.7 & 15.9 & 24.5 & \multicolumn{2}{c|}{26.3} & \multicolumn{3}{c}{21.0}  \\
	\end{tabular}
    \label{table:dataset}
\vspace{-0.2cm}
\end{table*}

Our proposed method is evaluated on two ensemble simulation datasets: Nyx~\cite{almgren2013nyx} and MPAS-Ocean~\cite{ringler2013multi}.
The details are shown in Table~\ref{table:dataset} (top) and detailed below. 

\textbf{Nyx}  \quad
Nyx is designed for cosmological simulations by Lawrence Berkeley National Laboratory.
Following scientists' suggestions, we study three parameters: the total matter density ($OmM \in [0.12, 0.55]$), the total density of baryons ($OmB \in [0.0215, 0.0235]$), and the Hubble constant ($h \in [0.55, 0.85]$). 
We randomly sampled 130 parameter settings from the parameter space and randomly picked 100 for training and 30 for testing. 
A 200-timestep cosmological simulation was conducted with each parameter setting, and a volume representing dark matter log density was generated. 
{The volume size is $512 \times 512 \times 512$, which means $W_0 = H_0 = L_0 = 512$. }
The three axes of the volume are $X$-axis, $Y$-axis, and $Z$-axis.

\textbf{MPAS-Ocean}  \quad
MPAS-Ocean is a simulation of the global ocean system developed by Los Alamos National Laboratory. 
Following scientists' suggestion, we studied four parameters: 
the amplitude of the ocean surface wind stress ($BwsA \in [0.0, 5.0]$), the critical bulk Richardson number (used to determine the strength of vertical mixing) ($CbrN \in [0.25, 1.00]$), the magnitude of the Gent McWilliams mesoscale eddy parameterization ($GM \in [600.0, 1500.0]$), and horizontal viscosity ($HV \in [100.0, 300.0]$). 
One hundred parameter settings were randomly sampled from the parameter space, among which we randomly picked 70 for training and 30 for testing. 
A 15-model-day ocean simulation was conducted with each parameter setting. 
Focusing on the eastern equatorial Pacific cold tongue, we extracted a region of interest (ROI) within $160^\circ W$ to $80^\circ E$, $26^\circ S$ to $26^\circ N$, and sea level to a depth of 200 meters, and generated volumes of size $1536 \times 768 \times 768$. 
The three axes of the volume are longitude ($\Theta$-axis), latitude ($\Phi$-axis), and depth ($D$-axis). 

\begin{figure}
  \centering
  \includegraphics[width=\linewidth]{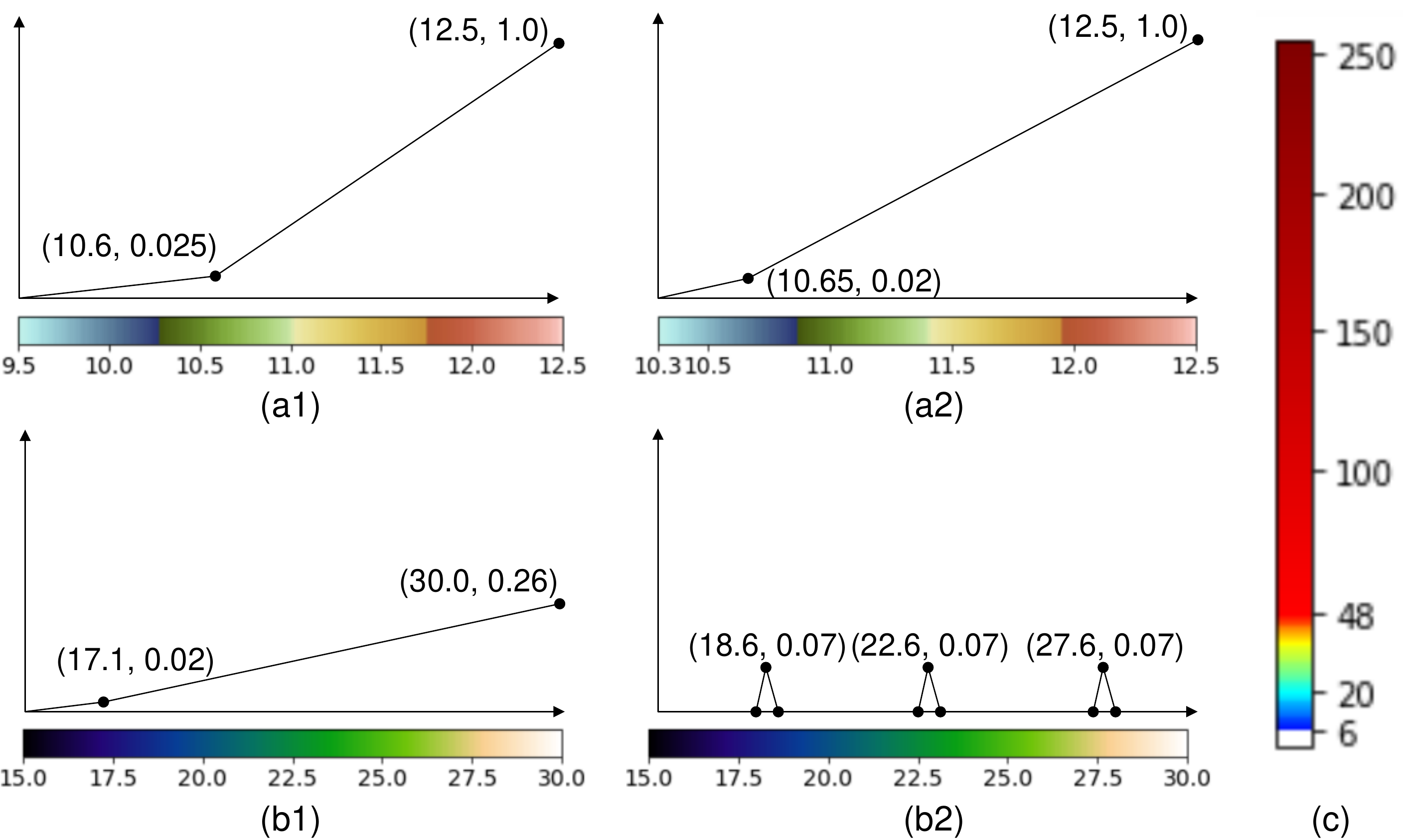}
  \caption{Transfer functions and color map. 
  (a) Transfer functions designed for Nyx. 
  (b) Transfer functions designed for MPAS-Ocean. 
  (c) Color map used for difference images.}
  \label{fig:tf}
\vspace{-0.4cm}
\end{figure}

\subsection{Implementation and Performance}
\label{subsection:implementation}

Our proposed method comprises two major components: (1) simulation runs and (2) VDL-Surrogate training and inference.  
The implementation details and performance analysis are reported below.  

\textbf{Simulation runs}  \quad 
We conducted our simulations on a supercomputer with 648 computation nodes. 
An Intel Xeon E5-2680 CPU with 28 cores and 128 GB memory is contained in a node.
We used 28 and 128 processes for Nyx and MPAS-Ocean simulations, respectively.
The simulation running time and output data size are reported in Table~\ref{table:dataset}. 
Among the 100 and 70 ensemble members in the Nyx and MPAS-Ocean datasets, 16 were selected for the RAE training. 

\begin{table} \footnotesize
\caption{{The size of view-dependent data after view-dependent sampling from three viewing directions for the MPAS-Ocean dataset. }}
    \centering
    \begin{tabular}{l|c|c|c} 
           & $W$ & $H$ & $L_0$   \\ \hline
         $\Theta$ & 384 & 384 & 1536  \\  \hline
         $\Phi$ & 384 & 768 & 768  \\ \hline
         $D$ & 768 & 384 & 768 \\ 
    \end{tabular}
    \label{table:vd_size_mpas}
\vspace{-0.4cm}
\end{table}

\textbf{VDL-Surrogate training and inference}  \quad
As explained in Section~\ref{section:overview}, VDL-Surrogate is composed of three RAEs and three VDL-Predictors, each pair for one axis.
For the Nyx dataset, the three axes are $X$, $Y$, $Z$,  {and after the view-dependent sampling, a volume with size $384 \times 384 \times 512$ is generated, i.e., $W = H = 384$. } 
For the MPAS-Ocean dataset, the three axes are $\Theta$, $\Phi$, $D$, {and Table~\ref{table:vd_size_mpas} provides the size of view-dependent data after performing view-dependent sampling from the three viewing directions.} 
VDL-Surrogate was implemented in PyTorch~\cite{paszke2019pytorch}.
The training and inferencing of an RAE or a VDL-Predictor were done on an NVIDIA Volta V100 GPU 16GB. 
Three RAEs or three VDL-Predictors are trained in parallel with three GPUs. 
During RAE training, we fixed the hyper-parameter $k_r = 64$ and set the batch size to fully utilize the 16GB GPU memory, where $k_r$ is the hyper-parameter used for controlling RAE's network size. 
The hyper-parameter $t$ controlling the latent representation's channel size was set to 3. 
During VDL-Predictor training, we set the batch size to 1 and set different $k_v$ for Nyx and MPAS-Ocean to fully use the 16GB GPU memory, where $k_v$ is the hyper-parameter used for controlling VDL-Predictor's network size. 
For both RAE and VDL-Predictor training, Adam is used as the optimizer with $\beta_1 = 0.0$ and $\beta_2 = 0.999$~\cite{kingma2014adam}.
The learning rate of RAE and VDL-Predictor was set to $5 \times 10^{-5}$. 

The network sizes of RAE and VDL-Predictor can be found in Table~\ref{table:dataset} (top, two rightmost columns).
We can see that RAE and VDL-Predictor's size sum is less than $4\%$ of the raw simulation data. 
The training and inference time of VDL-Surrogate is reported in Table~\ref{table:dataset} (bottom).
After RAEs are trained, the RAE encoding process takes around 10s and 26s for the Nyx and MPAS-Ocean datasets, respectively, so it would not significantly impact the simulations.
During inference, given a new simulation parameter setting, a VDL-Predictor forward pass and RAE decoding process take around 9s and 21s for the Nyx and MPAS-Ocean datasets, respectively.
After the final visualization, the image resolution for Nyx and MPAS-Ocean datasets is $600 \times 600$ and $800 \times 800$, respectively. 

\subsection{Evaluation Metrics}
\label{subsection:metric}

This section shows the data-level and image-level metrics used in our model evaluation.

\textbf{Data-level metrics} \quad
VDL-Surrogate predicts the simulation output so scientists can freely run visualization algorithms of their choice on the output data with desired algorithm parameters.  
Therefore, the quality of predicted simulation outputs is evaluated at the data level. 
Peak signal-to-noise ratio (PSNR) and normalized maximum difference (MD) are applied for measuring the grid-level difference and the error bound, respectively. 
{Note that in the following sections, for comparison between VDL-Surrogate and view-independent baseline methods, to perform a fair comparison, we first upsample the predicted view-dependent data back to full resolution via trilinear interpolation. }

\textbf{Image-level metrics}  \quad
We perform volume rendering {from $110$ viewpoints} with two different transfer functions for Nyx and two for the MPAS-Ocean dataset.
{We use Hierarchical Equal Area isoLatitude Pixelization  (HEALPix)~\cite{gorski2005healpix} to uniformly sample $110$ viewpoints on the viewing sphere, }and the transfer functions used in our experiments are provided in Figure~\ref{fig:tf}(a,b).
Moreover, inspired by previous works~\cite{han2019tsr, han2020v2v}, we give difference images to highlight the noticeable pixel differences with $\triangle \geq 6.0$ in the CIELUV color space).
The color map for the difference images is shown in Figure~\ref{fig:tf}(c).
Quantitatively, we apply structural similarity index measure (SSIM) and earth mover’s distance (EMD) between color histograms~\cite{berger2018generative} to evaluate the structural and distributional similarity between two volume rendered images.

\subsection{Motivation for Multiple View-Dependent Representations}
\label{subsection:multiple_view}

\begin{figure}
  \centering
  \includegraphics[width=0.6\linewidth]{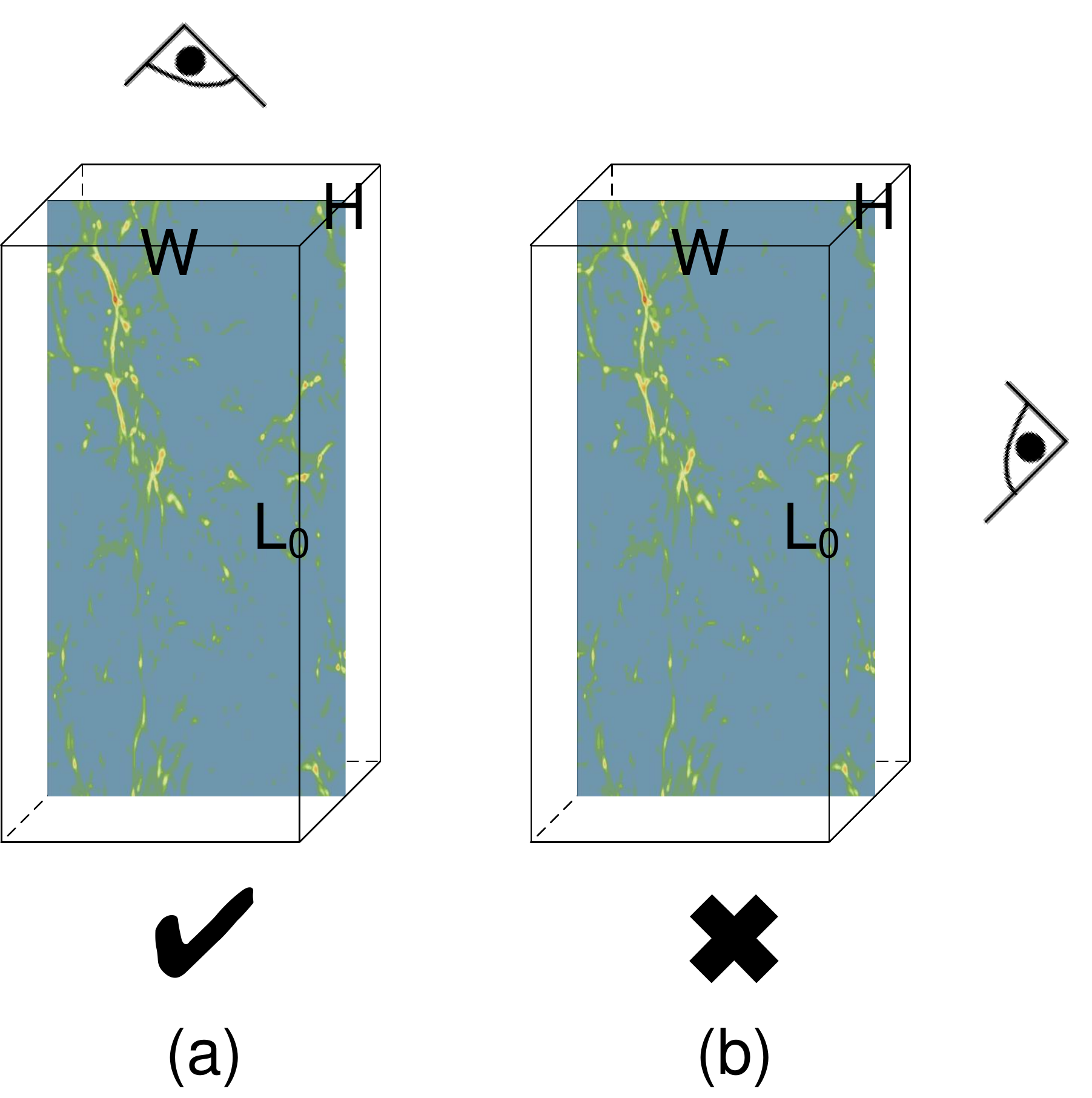}
  \caption{{The motivation for using view-dependent representations from three viewing directions.
  For the view-dependent data of size $W \times H \times L_0$, where the selected viewing direction is vertical, we prefer rendering from viewpoints close to the previous selected one (like (a)) rather than viewpoints far away (like (b)).}}
  \label{fig:vd_motivation}
\vspace{-0.6cm}
\end{figure}

{This section clarifies the motivation for using view-dependent representations from three viewing directions.
As mentioned in our workflow, one VDL-Predictor is trained to predict a view-dependent latent representation reduced in a selected viewing direction. 
In the inference stage, given the same viewing direction, we first predict the reduced view-dependent latent representation and decode the latent representations to the data space.
Because VDL-Predictor is trained based on the latent representations reduced at this selected viewing direction and we need an extra decoding process to predict the samples on the ray along the viewing direction, we do not recommend tiling these ray samples onto the screen.
As illustrated in Figure~\ref{fig:vd_motivation}, for the view-dependent data of size $W \times H \times L_0$, where the selected viewing direction is vertical, we prefer rendering from viewpoints close to the previous selected one (like (a)) rather than viewpoints far away (like (b)). }

\begin{table} \footnotesize
\caption{{Quantitative comparison of images generated with view-based interpolation (Interp) and predicted view-dependent data from three viewing directions (VDL$\Theta$, VDL$\Phi$ and VDL$D$) for the MPAS-Ocean dataset. } }
    \centering
    \begin{tabular}{ll|c|c|c|c} 
        ~ & ~ & Interp & VDL$\Theta$ & VDL$\Phi$ & VDL$D$ \\ \hline
        \multirow{2}{*}{TF1} &  SSIM  & \textbf{0.9888} & 0.9853 & 0.9878 & 0.9848  \\  
        ~ &  EMD  & 0.0021 & 0.0031  & 0.0030 & \textbf{0.0014}  \\ \cline{1-6} 
        \multirow{2}{*}{TF2} & SSIM & \textbf{0.9126} & 0.8855 & 0.8917 & 0.9079
        \\  
        ~  & EMD & \textbf{0.0012} & 0.0015 & 0.0013 & 0.0013  \\
    \end{tabular}
    \label{table:vd_motivation_mpas}
\end{table}

{We validate our method by evaluations on the MPAS-Ocean dataset.
In Table~\ref{table:vd_motivation_mpas}, we compare images generated with view-based interpolation and predicted view-dependent data from three viewing directions. 
The result shows that view-based interpolation increases SSIM under both TF1 and TF2 and decreases EMD under TF2 than using any single view-dependent data. }

\begin{figure*}
  \centering
  \includegraphics[width=0.88\linewidth]{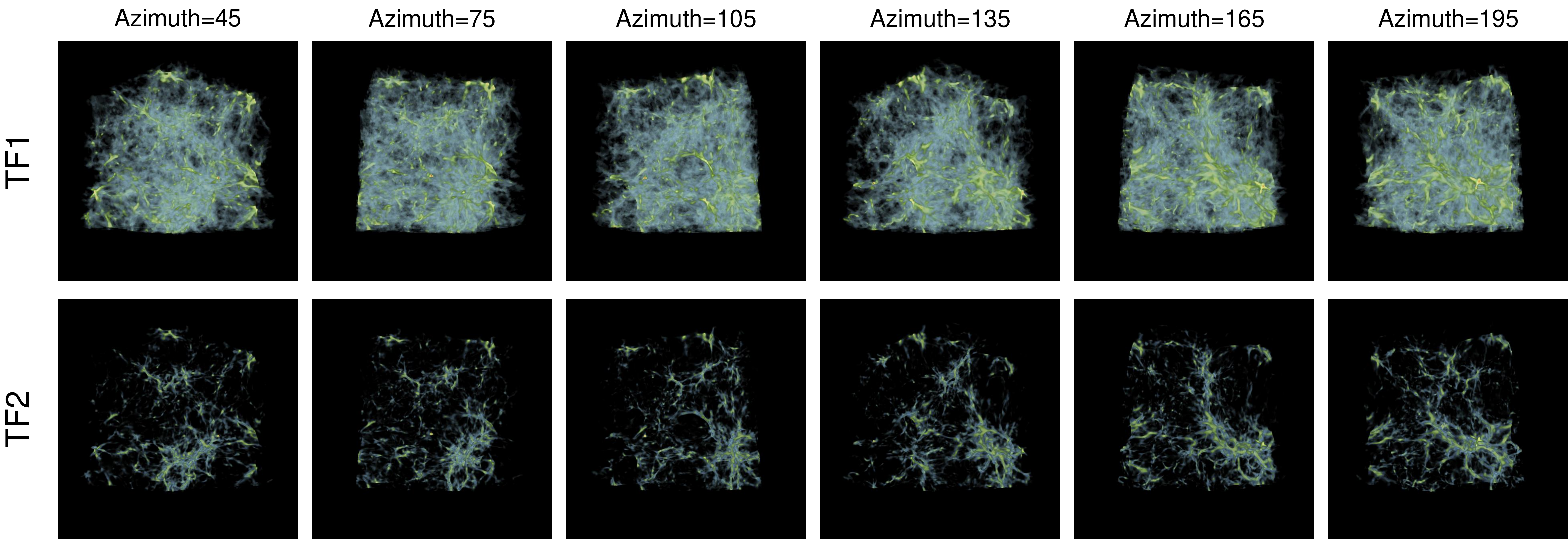}
  \caption{{Predicted final visualization images of the Nyx dataset for different transfer functions and varying viewpoints.}} 
  \label{fig:varying_vp_nyx}
\vspace{-0.2cm}
\end{figure*}

\begin{figure*}
  \centering
  \includegraphics[width=0.88\linewidth]{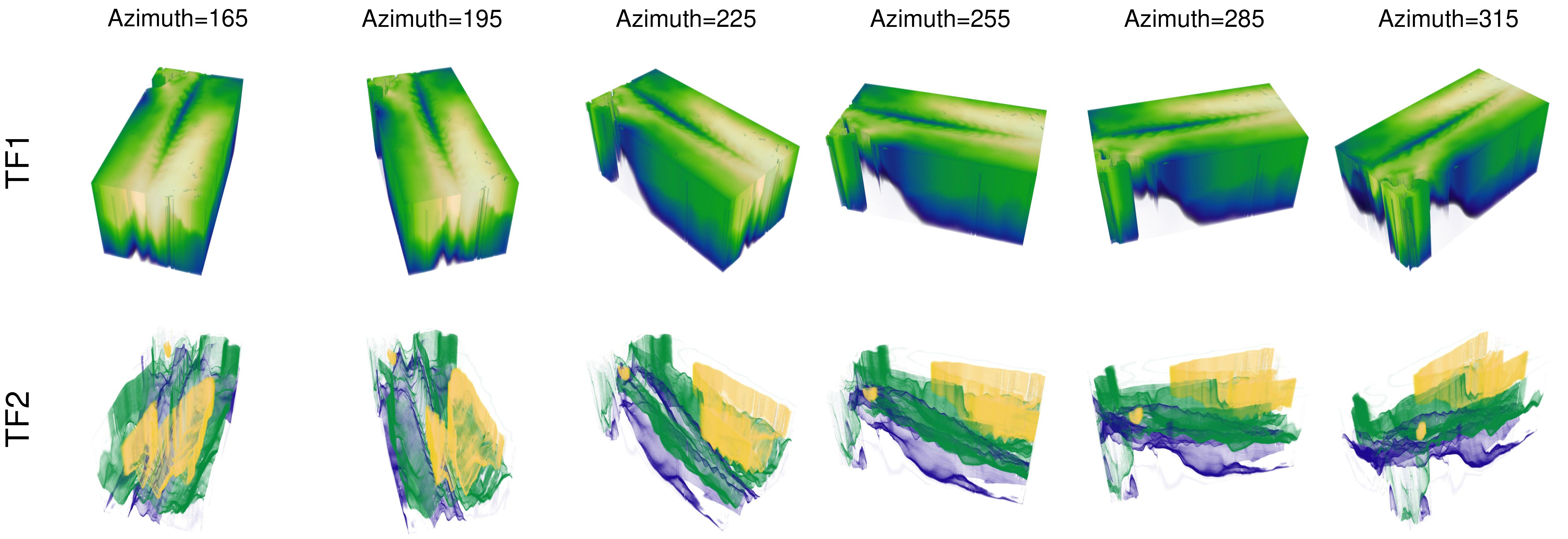}
  \caption{{Predicted final visualization images of the MPAS-Ocean dataset for different transfer functions and varying viewpoints.}}
  \label{fig:varying_vp_mpas}
\vspace{-0.4cm}
\end{figure*}

{Moreover, since our proposed surrogate utilizes view-dependent representations, it is vital to evaluate the performance of the surrogate with varying viewpoints. 
Figure~\ref{fig:varying_vp_nyx} and Figure~\ref{fig:varying_vp_mpas} provide predicted final visualization images for the Nyx and MPAS-Ocean datasets with different transfer functions and varying viewpoints, respectively.
More results can be found in our accompanying video.} 

\subsection{Comparison with Baseline Approaches}
\label{subsection:baseline}

\begin{table} \footnotesize
\caption{MPAS: quantitative evaluation of the inverse
distance weighting interpolation with different number of sampled data instances $g$. }
    \centering
    \begin{tabular}{l|c|c|c|c|c} 
         \# g & 1 & 2 & 3 & 4 & 5 \\  \hline
         PSNR (dB) & 38.18 & 37.87 & \textbf{39.08} & 38.58 & 38.63  \\  \hline
         MD & 0.1847 & 0.1715 & \textbf{0.1573} & 0.1626 & 0.1607 \\ 
    \end{tabular}
    \label{table:interp}
\end{table}

\textbf{Comparison with Interpolations} \quad
First, we compared our surrogate method with two alternative interpolation methods, including inverse distance weighting (IDW) interpolation and radial basis function (RBF) interpolation, to generate ensemble simulation data on MPAS-Ocean. 
IDW interpolation is one of the most widely used interpolation techniques for scientific data analysis~\cite{chen2012estimation}. 
It is straightforward to interpret and has low computational cost~\cite{lu2008adaptive}. 
In our experiment, We sampled a number of simulation outputs, denoted as $g$,  from the training dataset with the minimum Manhattan distance to the target test simulation output. 
We evaluated the number of samples $g$ from 1 to 5 at the data level, and the results are listed in Table~\ref{table:interp}. 
We present the results for $g=3$ since it obtained interpolation results with the highest PNSR and lowest MD. 
RBF interpolation is also widely used for scientific applications and is more complicated~\cite{wild2008orbit}.
In the experiment, Gaussian distribution is applied as the radial basis function, and backward propagation is used for optimization.

\begin{table} \footnotesize
\caption{{Quantitative comparison of the output predicted with VDL-Predictor, IDW interpolation, and RBF interpolation for the MPAS-Ocean dataset. }}
    \centering
    \begin{tabular}{l|c|c|c|c|c} 
           & VDL$\Theta$ & VDL$\Phi$ & VDL$D$ & IDW & RBF  \\ \hline
         PSNR (dB) &\textbf{42.14} & \textbf{43.20} & \textbf{41.04} & 39.08 & 33.49 \\  \hline
         MD & 0.5044 & 0.5008 & 0.6288 & 0.1573 & \textbf{0.1348} \\ 
    \end{tabular}
    \label{table:eval_interp_data_mpas}
\vspace{-0.4cm}
\end{table}

\begin{figure}
  \centering
  \includegraphics[width=0.6\linewidth]{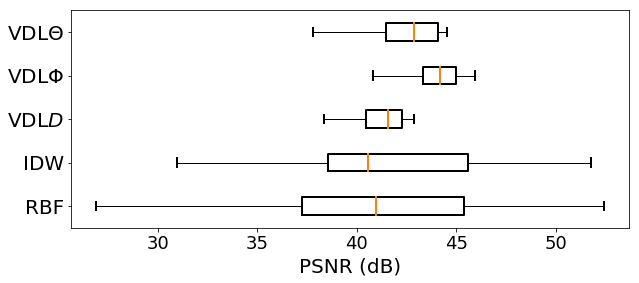}
  \caption{{Box plot comparing PSNR variants of VDL-Surrogate for three view directions, IDW interpolation, and RBF interpolation from different input simulation parameters forf the MPAS-Ocean dataset.}}
  \label{fig:boxplot_psnr}
\vspace{-0.4cm}
\end{figure}

\begin{table} \footnotesize
\caption{{Quantitative comparison of images generated with VDL-Surrogate, IDW interpolation, and RBF interpolation for the MPAS-Ocean dataset.} }
    \centering
    \begin{tabular}{ll|c|c|c} 
        ~ & ~ & VDLSurro & IDW & RBF \\ \hline
        \multirow{2}{*}{TF1} &  SSIM  & 0.9888 & \textbf{0.9959} & 0.9892 \\  
        ~ &  EMD  & 0.0021 &  \textbf{0.0008} & 0.0037  \\ \cline{1-5} 
        \multirow{2}{*}{TF2} & SSIM & 0.9126 & \textbf{0.9393}  & 0.9066
        \\  
        ~  & EMD & \textbf{0.0012} & \textbf{0.0012} & 0.0035 \\
    \end{tabular}
    \label{table:eval_interp_img}
\end{table}

\begin{figure}
  \centering
  \includegraphics[width=\linewidth]{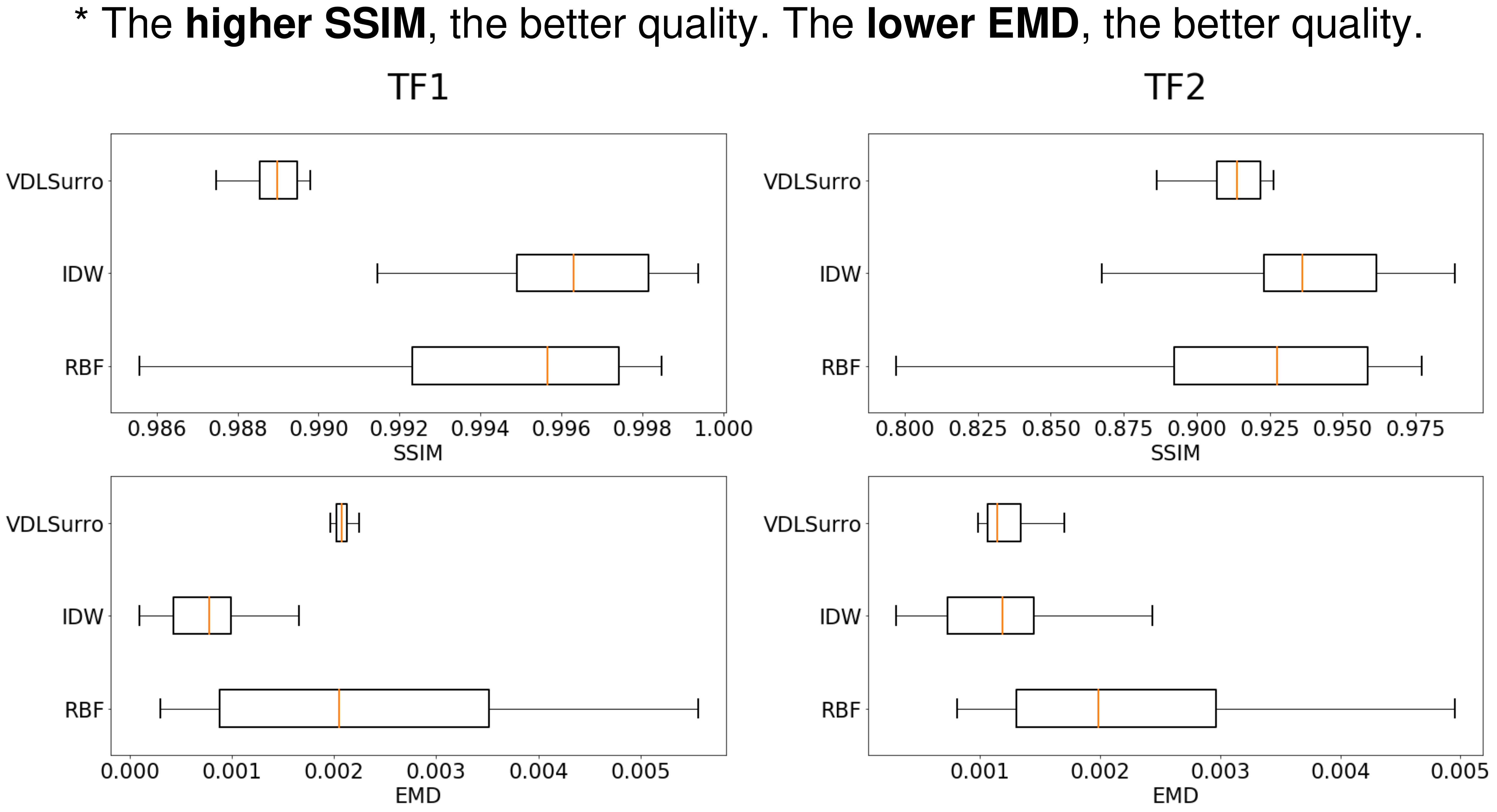}
  \caption{{Box plot comparing SSIM and EMD variants of VDL-Surrogate, IDW interpolation, and RBF interpolation from different input simulation parameters for the MPAS-Ocean dataset.}}
  \label{fig:boxplot_ssim_emd}
\vspace{-0.4cm}
\end{figure}

In Table~\ref{table:eval_interp_data_mpas}, we quantitatively compared VDL-Surrogate results against IDW and RBF interpolations at the data level (PSNR and MD). 
We found that our VDL-Surrogate produces higher PSNR than interpolations. 
VDL-Surrogate does not have a lower normalized maximum difference than interpolations, which can be explained because the loss function used in VDL-Surrogate training does not have a constraint on the error bound. 
To compare the PSNR variance of predicted simulations from different input simulation parameters, we render box plots, as shown in Figure~\ref{fig:boxplot_psnr}. 
Although IDW and RBF interpolation performed better for some input parameters, they have worse lower quartiles and medians, meaning they are less trustworthy. 

{Table~\ref{table:eval_interp_img} gives quantitative comparisons between the visualization images generated by VDL-Surrogate and IDW and RBF interpolations. 
Overall, VDL-Surrogate does not have higher SSIMs and lower EMDs compared with the simple IDW interpolation.
However, from Figure~\ref{fig:boxplot_ssim_emd}, we found that VDL-Surrogate has lower SSIM and EMD variance of final visualization images from different input simulation parameters than interpolations, which means the predictions from VDL-Surrogate are more stable among different ensemble members. 
For TF2, from the low minimum SSIM and high maximum EMD, we found that the IDW interpolation does not perform well for some ensemble members. 
For TF1, the IDW interpolation has extremely high SSIM and low EMD, maybe because TF1 is a high-opacity transfer function and IDW interpolation performs well on the volume surface. 
Another benefit of using surrogate models is that the prediction only requires trained models smaller than 1GB rather than the 65GB raw simulation outputs. }

\begin{table} \footnotesize
\caption{Quantitative comparison of images generated with VDL-Surrogate and InSituNet. }
    \centering
    \begin{tabular}{lll|c|c} 
       &  &  & VDLSurro & InSituNet \\ \hline
       \multirow{4}{*}{Nyx} & \multirow{2}{*}{TF1} &  SSIM  & \textbf{0.9291} & 0.7269 \\  
        ~ & ~ &  EMD  & 0.0068 &  \textbf{0.0023}  \\ \cline{2-5} 
        ~ & \multirow{2}{*}{TF2} & SSIM & \textbf{0.9634} & 0.8345 
        \\  
        ~ & ~  & EMD & 0.0012 & \textbf{0.0007} \\ \hline
       \multirow{4}{*}{MPAS-Ocean} & \multirow{2}{*}{TF1} &  SSIM & \textbf{0.9888} & 0.9540 \\  
        ~ & ~ &  EMD  & \textbf{0.0021} & 0.0025  \\ \cline{2-5} 
        ~ & \multirow{2}{*}{TF2} & SSIM & \textbf{0.9126} & 0.8423 \\  
        ~ & ~  & EMD & \textbf{0.0012} & 0.0021 \\ 
    \end{tabular}
    \label{table:eval_insitunet_img}
\vspace{-0.4cm}
\end{table}

\textbf{Comparison with InSituNet} \quad
We compared our approach with the state-of-the-art image-based method InSituNet~\cite{he2019insitunet} for image-based analyses on both Nyx and MPAS-Ocean datasets.
As an image-based method, InSituNet does not support data-level comparison, so it is only used for the comparison of image-based analyses. 
Note that we need to train multiple InSituNets if we want to apply different visual mappings. 
 
\begin{figure}
  \centering
  \includegraphics[width=\linewidth]{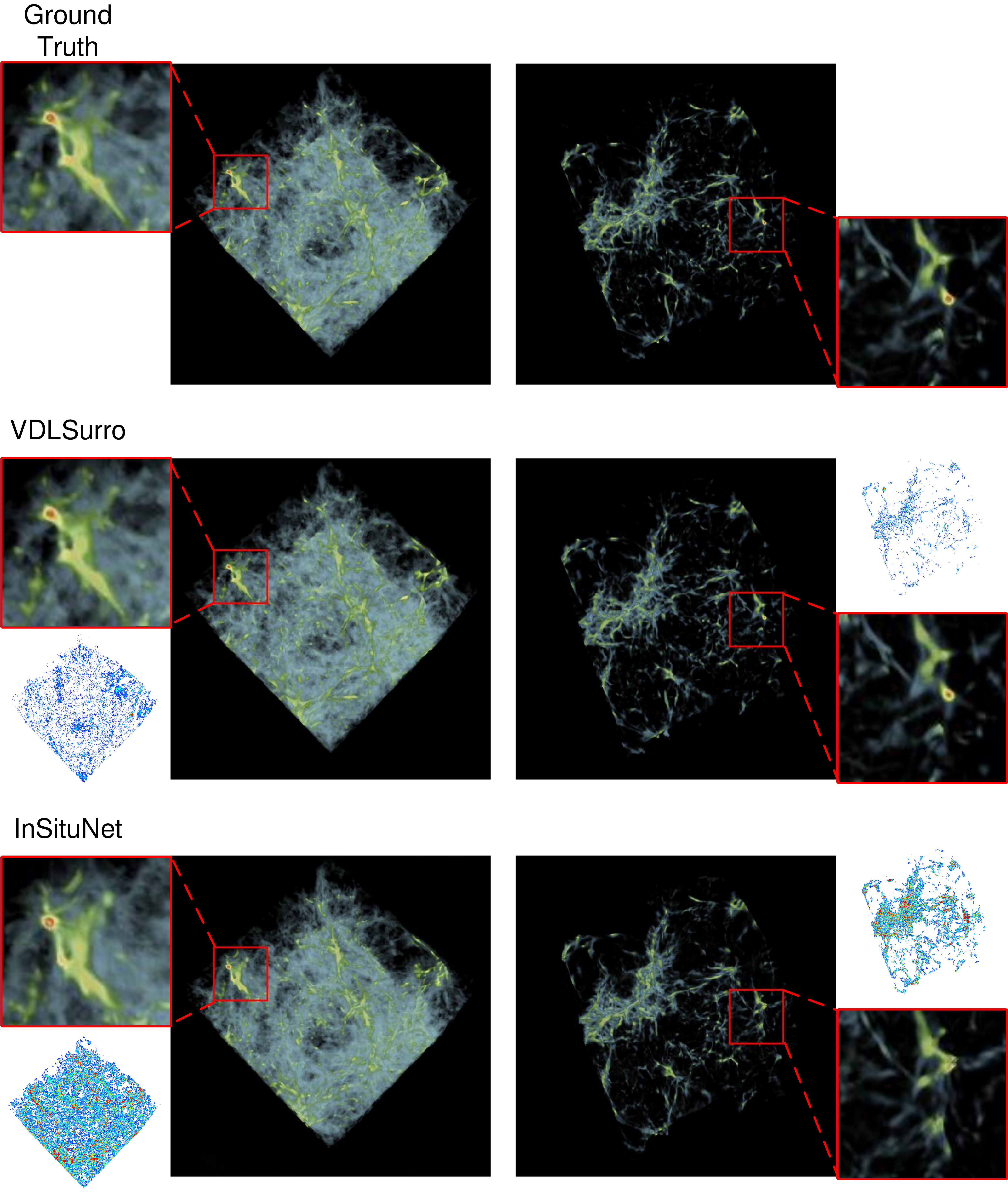}
  \caption{Comparison of the images generated using VDL-Surrogate and InSituNet for the Nyx dataset with the ground truth images.}
  \label{fig:nyx_img_cmp}
\vspace{-0.4cm}
\end{figure}

\begin{figure}
  \centering
  \includegraphics[width=\linewidth]{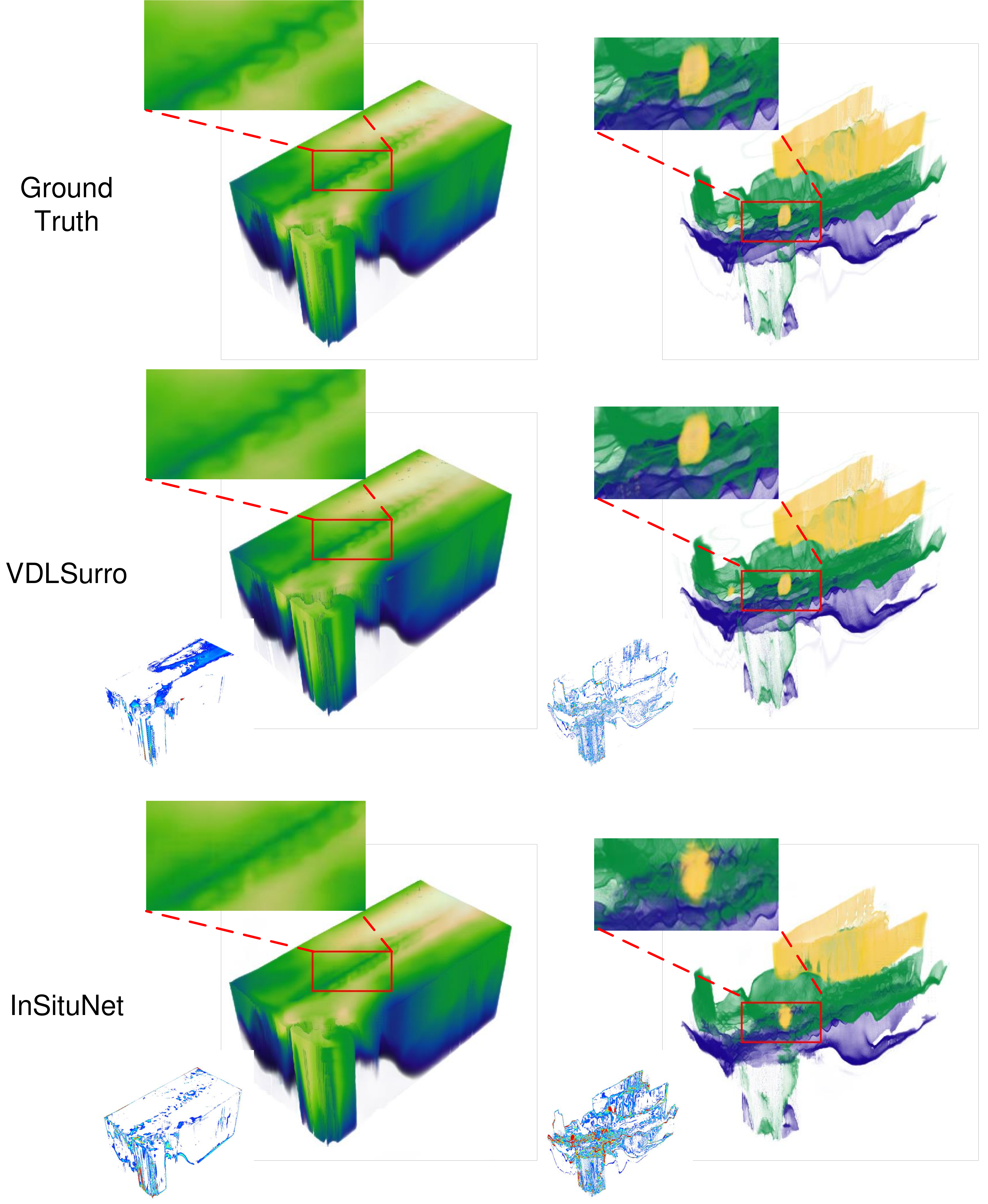}
  \caption{Comparison of the images generated using VDL-Surrogate and InSituNet for the MPAS-Ocean dataset with the ground truth images.}
  \label{fig:mpas_img_cmp}
\vspace{-0.4cm}
\end{figure}

Table~\ref{table:eval_insitunet_img} shows the image-level quantitative results on both Nyx and MPAS-Ocean datasets.
Compared with InSituNet, VDL-Surrogate produces images with higher SSIM 
For the Nyx data, the images generated by VDL-Surrogate are with higher EMD. 
We think it is because VDL-Surrogate does not directly predict images, which causes the image color distribution similarity not to be as good as InSituNet. 
The qualitative comparison between images generated by VDL-Surrogate and InSituNet for the Nyx dataset with ground truth images can be found in Figure~\ref{fig:nyx_img_cmp}. 
VDL-Surrogate generated predicted images with more accurate details, while InSituNet failed to recover these important details.
For example, as shown in the zoom-in patch, VDL-Surrogate correctly predicted the red spot with a high dark matter density, while InSituNet provided a wrong spot (left) or missed it (right).    
In Figure~\ref{fig:mpas_img_cmp}, we compared the images generated by VDL-Surrogate and InSituNet for the MPAS-Ocean dataset with ground truth images.
The first column shows images rendered with a high-opacity transfer function, which helps us analyze the sea level and vertical cross-section ocean temperature. 
The most important phenomenon in this ocean region is the equatorial cold tongue reflected by the blue stripe on the sea level. 
Our VDL-Surrogate accurately reflects the temperature, while the image provided by InSituNet is blurry, which cannot give enough details for oceanographers to have an effective preview and a meaningful analysis of the cold tongue. 
Images in the second column show the effect of blending three different isothermal surfaces. 
Our VDL-Surrogate gives a high-quality prediction, while the prediction by InSituNet is blurry, and oceanographers cannot know the relationship between the three isothermal surfaces in this region.

\subsection{Case Study: Parameter Space Exploration}
\label{subsection:case}

This section provides two case studies on the Nyx and MPAS-Ocean ensemble simulations, showing that scientists can use our proposed VDL-Surrogate to have an efficient parameter space exploration.  

\subsubsection{Case Study with the Nyx Simulation}
\begin{figure}
  \centering
  \includegraphics[width=\linewidth]{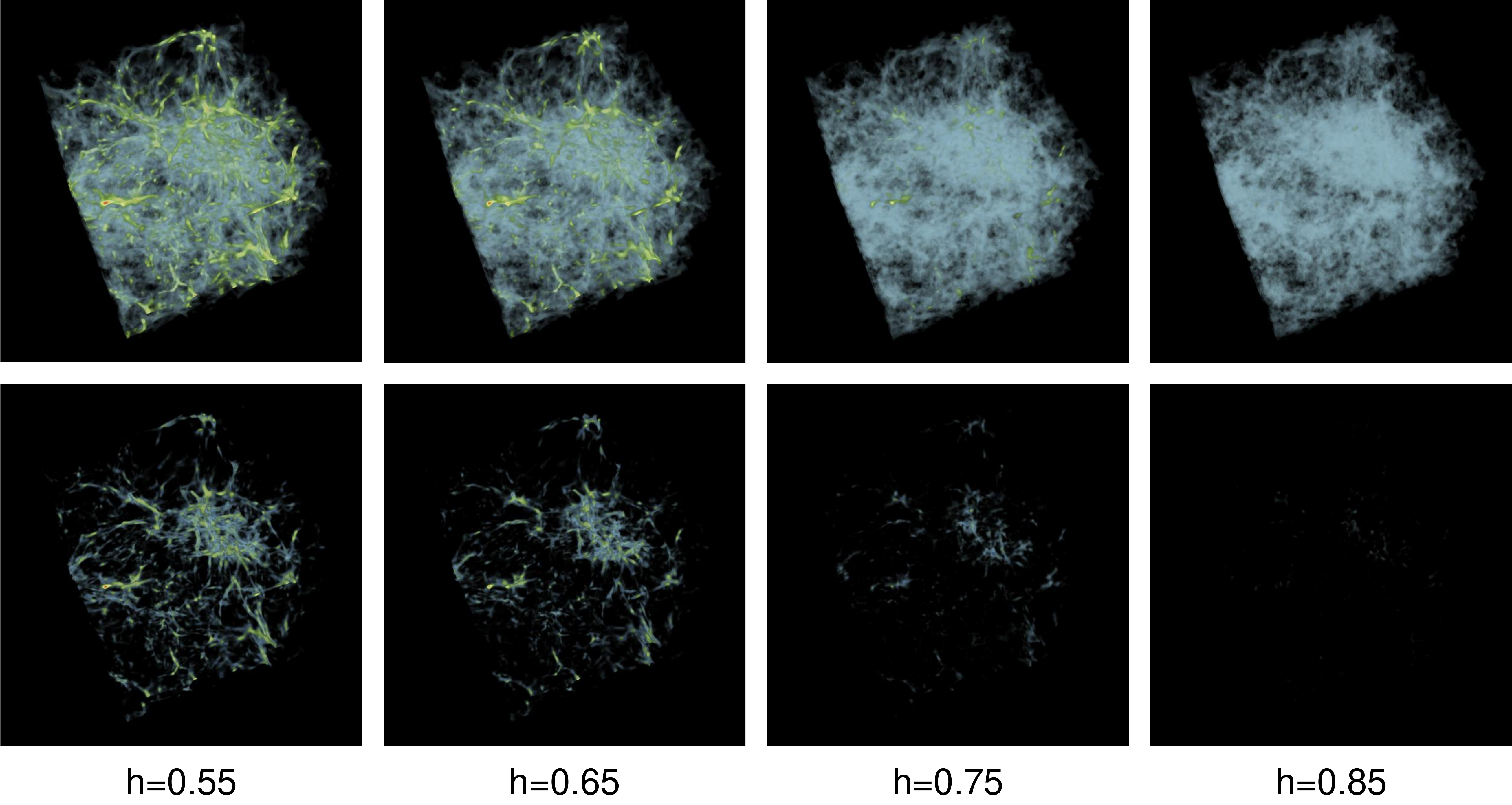}
  \caption{Case study for the predicted images rendered with two different transfer functions using four different $h$ values.}
  \label{fig:nyx_case}
\vspace{-0.4cm}
\end{figure}

\begin{figure*}
  \centering
  \includegraphics[width=\linewidth]{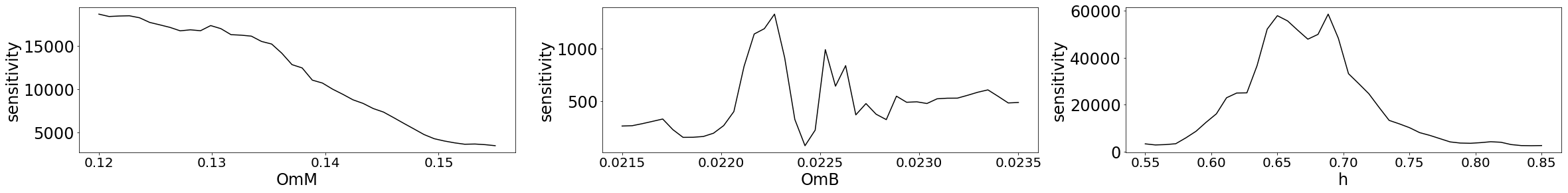}
  \caption{The line graph visualization of different simulation parameters' sensitivity for the Nyx dataset.}
  \label{fig:nyx_sensitivity}
\vspace{-0.4cm}
\end{figure*}

In the first case study, the three simulation parameters (i.e., $OmM$, $OmB$, and $h$) on the Nyx dataset are analyzed.
First, we calculate the gradients of the three parameters to the predicted simulation output as the parameters' sensitivity. 
The results are illustrated in Figure~\ref{fig:nyx_sensitivity}.
Considering the four plots' scale, We found that $h$ is the most sensitive among the three simulation parameters, $OmM$ is the second most sensitive, and $OmB$ is the least sensitive. 

Second, we focused on the most sensitive parameter, $h$, and analyzed how it influences the density field. 
We fixed $OmM = 0.1375$, $OmB = 0.0225$, and sampled $h$ from the set $\{0.55, 0.65, 0.75, 0.85\}$. 
We provided volume-rendered images with two different functions, as shown in Figure~\ref{fig:nyx_case}. 
Scientists can visualize the high-density and low-density regions together with the first transfer function or only look at the high-density region with the second transfer function.
From the images, we found that the high-density region becomes smaller as the Hubble Constant $h$ increases. 
{More explorations can be found in our accompanying video.}

\subsubsection{Case Study with the MPAS-Ocean Simulation}
\begin{figure}
  \centering
  \includegraphics[width=\linewidth]{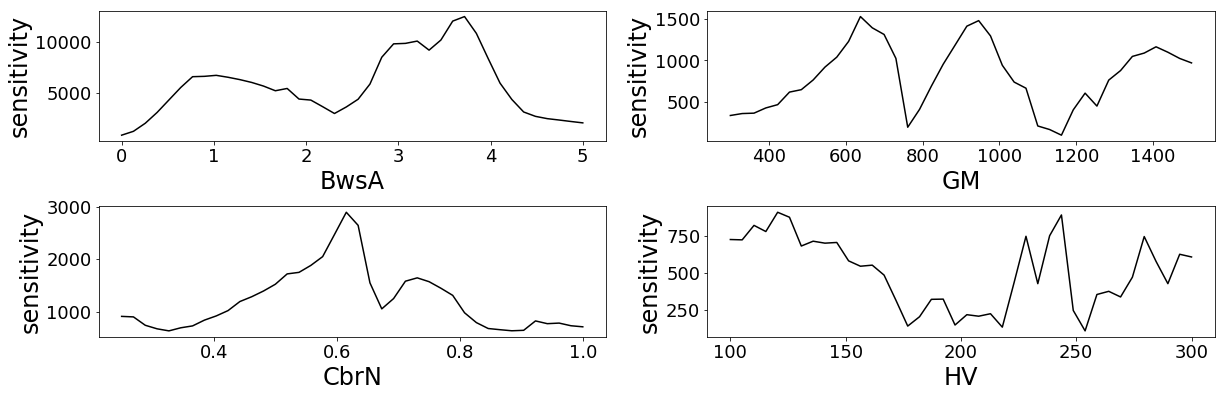}
  \caption{The line graph visualization of different simulation parameters' sensitivity for the MPAS-Ocean dataset.}
  \label{fig:mpas_sensitivity}
\end{figure}

Our second case study focuses on the impact of different simulation parameters (i.e., $BwsA$, $CbrN$, $GM$, and $HV$) on the MPAS-Ocean simulation. 
First, to probe the sensitivity of the four parameters, as illustrated in the line graphs in Figure~\ref{fig:mpas_sensitivity}, we calculate the gradients of the parameters to the predicted simulation output. 
The rank of the variable sensitivity is $BwsA > CbrN > GM > HV$.   

\begin{figure}
  \centering
  \includegraphics[width=\linewidth]{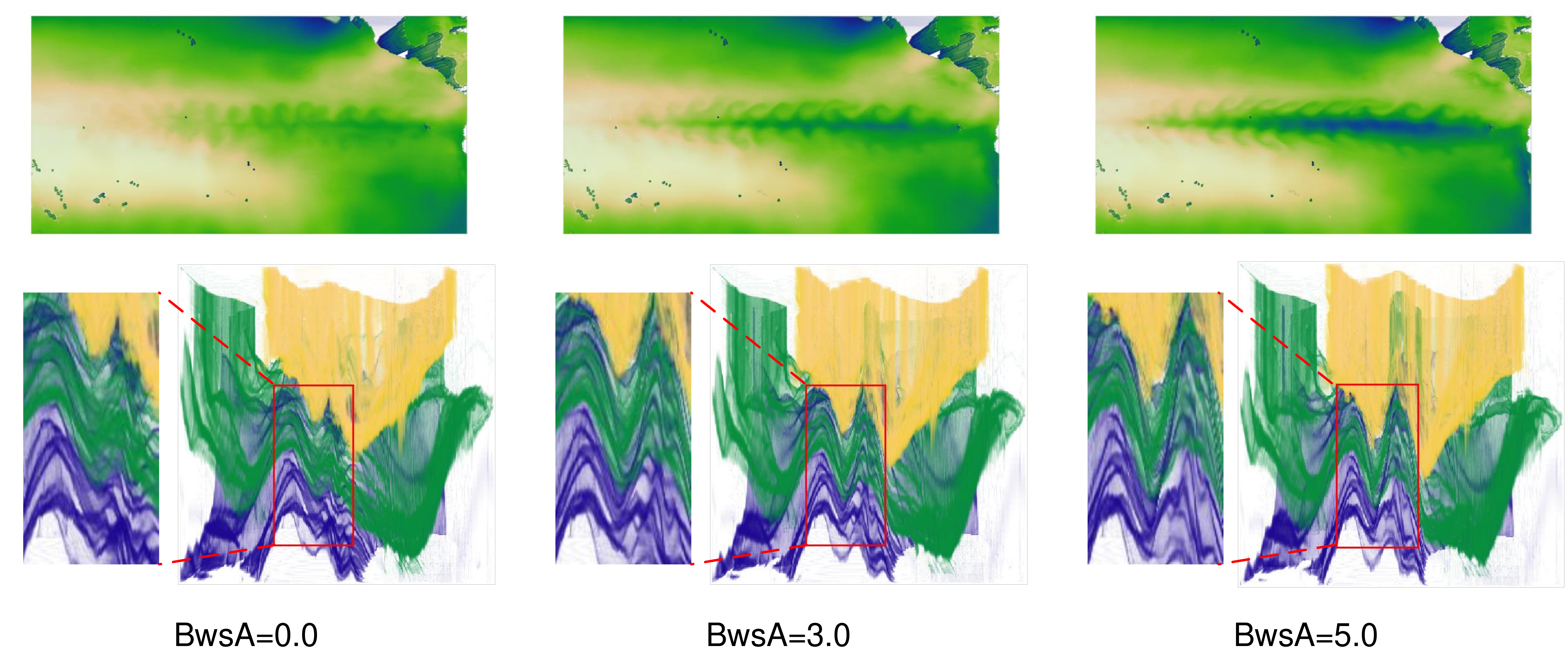}
  \caption{Case study for the predicted images rendered with two different transfer functions using four different $BwsA$ values.}
  \label{fig:mpas_case}
\vspace{-0.4cm}
\end{figure}

Second, we display how the most sensitive simulation parameter, $BwsA$, influences the ocean temperature.
We fixed $GM = 900.0$, $CbrN = 0.625$, $HV =200.0$, and chose 3 different $BwsA$ values: 0.0, 3.0, and 5.0.  
To begin with, we picked the first transfer function designed for MPAS-Ocean shown in Figure~\ref{fig:tf} and selected the top viewpoint to visualize the sea level temperature map. 
As shown by the top row in Figure~\ref{fig:mpas_case}, when the amplitude wind stress increases, the blue stripe in the image becomes more evident, meaning there is a stronger equatorial cold tongue in the eastern Pacific. 
One reason causing the phenomenon is the up-welling of cold subsurface water along the equator. 
To illustrate that, we used the second transfer function designed for MPAS-Ocean shown in Figure~\ref{fig:tf} and picked one side viewpoint to visualize three isothermal surfaces. 
From the bottom row in Figure~\ref{fig:mpas_case}, we can see that the surfaces have a higher curvature when $BwsA$ increases, which means the colder water up-welling effect is more significant. 
{Please see accompanying video for additional results.}

\section{Discussion and Limitations}
This section discusses the benefit of our proposed approach compared with another view-dependent surrogate model InSituNet~\cite{he2019insitunet} and discusses the limitations of our approach. 

We demonstrate that our work handles the three limitations mentioned in the InSituNet paper: (1) low resolution of output images; (2) low accuracy of output images; (3) poor flexibility when exploring different visual mapping parameters.
The key to our success is that VDL-Surrogate only takes simulation parameters as input and outputs the corresponding representation without encoding visual mapping and viewpoint parameters. 
First, VDL-Surrogate supports resolutions of predicted images to $600 \times 600$ and $800 \times 800$, respectively, for the Nyx and MPAS-Ocean datasets, while InSituNet limits the resolutions to only $256 \times 256$.
Moreover, our GPU memory usage is smaller than InSituNet (48GB vs. 320GB).
Without encoding viewpoint parameters, the number of training instances is only proportional to the number of ensemble members, not the number of ensemble members times the number of selected viewpoints.
Therefore, we do not need to use low resolutions to support large batch sizes when only a limited amount of GPU memory is available.
Second, VDL-Surrogate produces results with much higher accuracy than InSituNet, which is validated by our experiments. 
The key is not letting networks encode viewpoint parameters so as to  drastically reduces the variance of the training instances and  significantly mitigate the training difficulty.
Last, scientists can use their desired visual mappings to perform parameter space exploration, as illustrated in our experiment. 
We do not try to cover the large joint space of all possible simulation and visualization parameters. 
Instead, our training goal is to produce representations in the data space, allowing us to apply different visual mappings. 

One limitation of our work is that we only select view directions parallel to the three main axes. 
Theoretically, we can choose arbitrary viewpoints to perform ray casting and encode samples along the ray. 
However, if the view direction is not parallel to the main axes, the encoded representation is not in a cuboid.
CNNs can not directly handle the representation unless we consider its bounding box, which leads to unnecessary memory waste.
In the future, we would like to treat the representation as a graph and use graph neural networks to process it directly. 

Like many other machine learning works, another limitation of our approach is the long offline training time, which is around 50 hours for RAE and VDL-Predictor.
In the future, we plan to take advantage of high-performance machine learning techniques and speed up the network training with more GPUs by using the data-parallel technique provided by PyTorch~\cite{paszke2019pytorch}. 

\section{Conclusions}

In conclusion, we propose VDL-Surrogate, a view-dependent neural-network-latent-based surrogate model for parameter space exploration of ensemble simulations that allows high-resolution visualizations and user-specified visual mappings.
We use view-dependent latent representations to replace the raw data as the input and output of the visualization surrogate to train the surrogate with limited GPU memory. 
In the inference stage, given a new input simulation parameter, we first predict the latent representation and then {decode the representation to data space}.
Scientists can apply different user-specified visual mappings to obtain different visualization results. 
We demonstrate the effectiveness and efficiency of VDL-surrogate by comprehensive quantitative and qualitative evaluations on cosmological and ocean ensemble simulation datasets.

\acknowledgments{
This work is supported in part by the US Department of Energy SciDAC program DE-SC0021360, National Science Foundation Division of Information and Intelligent Systems IIS-1955764, and National Science Foundation Office of Advanced Cyberinfrastructure OAC-2112606.}

\bibliographystyle{abbrv-doi}

\bibliography{reference}
\end{document}